\documentclass{aa}
\usepackage{graphicx}
\usepackage{txfonts}

\begin{document}

\title{\bf Dust formation in the ejecta of the Type II-P supernova 2004dj}

\author{
T. Szalai
\inst{1} 
\and
J. Vink\'o
\inst{1} 
\and
Z. Balog
\inst{2}
\and
A. G\'asp\'ar
\inst{3}
\and
M. Block
\inst{3}
\and
L L. Kiss
\inst{4,5}
}

\institute{
Department of Optics and Quantum Electronics, University of
Szeged, D\'om t\'er 9., Szeged H-6720, Hungary\\
\email{szaszi@titan.physx.u-szeged.hu; vinko@titan.physx.u-szeged.hu}
\and
Max-Planck-Institut f\"ur Astronomie, K\H{o}nigstuhl 17, D-69117 Heidelberg, 
Germany
\and
Steward Observatory, University of Arizona, Tucson, AZ 85721, USA
\and
Konkoly Observatory of the Hungarian Academy of Sciences, H-1525 Budapest, P.O. Box 67, Hungary
\and
Sydney Institute for Astronomy, School of Physics A28, University of Sydney, NSW 2006, Australia}

\date{Received ..; accepted ..}

\abstract{}{}{}{}{}

\abstract
{}
{Core-collapse supernovae (CC SNe), especially Type II-Plateau ones, are thought to be important contributors to
cosmic dust production. SN 2004dj, one of the closest and brightest SN since 1987A, offered a good opportunity to examine
dust-formation processes. To find signs of newly formed dust, we analyze all available mid-infrared (MIR) 
archival data from the {\it Spitzer Space Telescope}.}
{We re-reduced and analyzed data from IRAC, MIPS, and IRS instruments obtained between +98 and
+1381 days after explosion and generated light curves and spectra for each epoch. Observed spectral energy distributions 
are fitted with both analytic and numerical models, using the radiative-transfer 
code MOCASSIN for the latter ones. We also use imaging polarimetric data obtained at +425 days by the {\it Hubble Space Telescope}.}
{We present convincing evidence of dust formation in the ejecta of SN 2004dj from MIR light curves and spectra.
Significant MIR excess flux is detected in all bands between 3.6 and 24 $\mu$m. 
In the optical, a $\sim$0.8~\% polarization is also detected at a 2-sigma level, which exceeds the interstellar polarization
in that direction. Our analysis shows that the freshly-formed dust around SN 2004dj 
can be modeled assuming a nearly spherical shell that contains amorphous carbon grains,
which cool from $\sim$700 K to $\sim$400 K between +267 and +1246 days. 
Persistent excess flux is found above 10 $\mu$m, 
which is explained by a cold ($\sim$115 K) dust component. If this cold dust is of circumstellar origin, it is likely
to be condensed in a cool, dense shell between the forward and reverse shocks. 
Pre-existing circumstellar dust is less likely, but cannot be ruled out. 
An upper limit of $\sim$8 $\times$ 10$^{-4} M_{\odot}$ is derived for the dust mass, which is similar to  
previously published values for other dust-producing SNe. }
{}


\maketitle

\section{Introduction}\label{intro}

Core-collapse supernovae (CC SNe) are usually interpreted as the endpoints in the life cycle of massive stars 
(M $\gtrsim$ 8 $M_{\odot}$; e.g. Woosley et al. 2002). 
While it is widely accepted that these high-energy events have significant 
effects on the birth and evolution of other stars, there has been a long-standing argument whether they 
also play an important role in the formation of interstellar dust.
 
Theories on the dust-production of CC SNe have been around for over 40 years (Cernuschi et al. 1967;
Hoyle \& Wickramasinghe 1970). These early hypotheses were later supported by studies of isotopic anomalies in meteorites 
(Clayton 1979; Clayton et al. 1997; Clayton \& Nittler 2004). Additionally, several papers were published 
about the unexpectedly large dust 
content of high-redshift galaxies (Pei et al. 1991; Pettini et al. 1997; Bertoldi et al. 2003), which suggested that CC SNe
could be the main sources of interstellar dust in the early (and maybe in the present) Universe 
(Todini \& Ferrara 2001; Nozawa et al. 2003; Morgan \& Edmunds 2003). The average lifetimes of CC SNe 
are short enough to produce dust at high redshifts, i.e. during $\sim$1 Gyr, thus these SNe are
better candidates for explaining the presence of early dust than other possible sources, 
like AGB stars (e.g. Dwek et al. 2007). 

Although these theories have not been
disclaimed so far, there are several other possible mechanisms of dust formation in distant 
galaxies. Valiante et al. (2009) showed that different star-formation histories of galaxies should be taken into account to 
avoid underestimating the contribution of AGB stars to the whole dust production. Another possibility is the 
condensation of dust grains in quasar winds (Elvis et al. 2002), which has also been supported by observations 
(Markwick-Kemper et al. 2007). On the other hand, there are many high-z galaxies where SN-explosions are the only viable 
explanation for the observed amount of dust (Maiolino et al. 2004; Stratta et al. 2007; Michalowski et al. 2010).
Detailed studies by Matsuura et al. (2009) suggest a 'missing dust-mass problem' in the Large Magellanic
Cloud (LMC), similar to the situation in distant galaxies described above. 

The two main problems with the observable dust content around SNe are the amount and the origin. Models of Kozasa et
al. (1989), Todini \& Ferrara (2001), and Nozawa et al. (2003) predict 0.1-1 $M_{\odot}$ of newly formed dust after a 
CC SN event. More recently, Bianchi \& Schneider (2007), Kozasa et al. (2009), and Silvia et al. (2010) have found 
similar values via numerical modeling of the survival of dust grains. They have found that the dust amount 
depends on the type of SN and also the density of the local ISM, which must be taken into account when estimating the 
contribution of CC SNe to the dust content in high-redshift galaxies.
 
Despite the nearly concordant results of different models, direct observations have not confirmed the massive
dust production by CC SNe yet (although a detailed analysis was possible only in a few cases). 
The first evidence for dust condensation was observed in SN~1987A (e.g. Danziger et al. 1989, 1991; Lucy et al. 1989, 1991; 
Roche et al. 1993 and Wooden et al. 1993; revisited by Ercolano et al. 2007). The observed signs of dust formation 
were $i)$ a strong decrease of optical fluxes around 500 days after explosion, $ii)$ an increase of mid-infrared (MIR) fluxes 
at the same time, and $iii)$ increasing blueshift and asymmetry of the optical emission lines because of an attenuation
of the back side of the ejecta by the freshly synthesized dust. The mass of
newly condensed dust was estimated to be $\sim$ 10$^{-4}$ $M_{\odot}$.
Similar effects were also detected in the ejecta of Type II-P SN~1999em and Type Ib/c SN~1990I (Elmhamdi et al. 2003, 2004).

The launch of the {\it Spitzer Space Telescope} (hereafter {\it Spitzer}) in 2003 and {\it AKARI} in 2006 
provided additional opportunities to observe dust around CC SNe by following the evolution of MIR light curves and spectra.
Evidence for newly condensed dust were obtained for SN~2003gd (Sugerman et al. 2006; Meikle et al. 2007), 2004et (Kotak et
al. 2009), 2007od (Andrews et al. 2010), and Type Ib/c SN~2006jc (Nozawa et al. 2008; Mattila et al. 2008; 
Tominaga et al. 2008 and Sakon et al. 2009). 
In every case, the estimated mass of recently formed dust was between $10^{-5}$ - $10^{-3}$ $M_{\odot}$.  
Note that for SN~2003gd Sugerman et al. (2006) calculated a value of 0.02 $M_{\odot}$, but their result was questioned 
by Meikle et al. (2007). These values are significantly lower than the theoretically predicted ones and 
tend not to support an SN origin for observable amounts of dust in the local and distant universe. 

There are other ways to discover dust around SNe. As it was found in some CC SNe 
(e.g. SN~1998S, Pozzo et al. 2004; SN~2005ip, Smith et al. 2009 and Fox et al. 2009; SN~2006jc, Smith et al. 2008a; 
SN~2006tf, Smith et al. 2008b; SN~2007od, Andrews et al. 2010), 
dust grains may condense in a cool dense shell (CDS) that is generated between the forward and reverse shock waves during 
the interaction of the SN ejecta and the pre-existing circumstellar medium (CSM). The CDS may affect both the light curves 
and the spectral line profiles.  
Another possibility is the thermal radiation of pre-existing dust in the CSM that is re-heated by the SN as an
IR-echo (Bode \& Evans 1980; Dwek 1983, 1985; Sugerman 2003), which can be observed as an infrared excess 
(e.g. SN~1998S, Gerardy et al. 2002, Pozzo et al. 2004; SN~2002hh, Barlow et al. 2005, Meikle et al. 2006; 
SN~2006jc, Mattila et al. 2008; SN~2004et, Kotak et al. 2009; SN~2008S, Botticella et al. 2009). 
These results seem to support the hypothesis that pre-explosion mass-loss processes of progenitor stars 
could play a more important role
in dust formation than the explosions of CC SNe (see also Prieto et al. 2008 and Wesson et al. 2010).

Recent studies of SN remnants (SNRs) could not solve the question of the amount of dust produced by SNe. 
By obtaining far-IR and sub-mm observations, many groups estimated the amount 
of newly condensed dust grains in Cas~A (Dunne et al. 2003; Krause et al. 2004; Rho et al. 2008). Their results varied 
between $0.02$ and $2$ $M_{\odot}$, while the values for Kepler SNR showed an even larger discrepancy 
(0.1-3 $M_{\odot}$ by Morgan et al. 2003, and $5\times 10^{-4}$ $M_{\odot}$ by Blair et al. 2007). 
Stanimirovic et al. (2005) and 
Sandstrom et al. (2009) studied the MIR data of SNR~1E0102.2-7219 in the Small Magellanic Cloud (SMC) and found 
evidence for $\sim 1-3 \times 10^{-3}$ $M_{\odot}$ of dust formed in the ejecta (the authors of the latter paper also 
calculated that the amount of cold dust that is observable only at longer wavelengths could be even higher
by two orders of magnitude). 
 
Estimating the mass of dust around SNe is complicated, and the results are strongly
model-dependent. Several ideas have been proposed to explain the discrepancies between
observations and theory. Using clumpy grain-density distributions (Sugerman et al. 2006, Ercolano et al. 2007) 
leads to a dust mass in a SN ejecta at least one order of magnitude higher than by assuming a smooth density distribution.  
Regarding dust enrichment in high-z galaxies, the low dust production rates of observed CC SNe might be compensated 
by using a top-heavy initial mass function (IMF), resulting in more SNe per unit stellar mass (Michalowski et al. 2010 
and references therein). 
It is also a possibility that SN explosions provide only the dust seeds, while the bulk of the dust mass is accumulated 
during grain growth in the ISM (Draine 2003, 2009; Michalowski et al. 2010 and references therein). It has also been noted
that estimates of the dust content at high redshifts should be considered with caution because of 
the uncertainties in interpreting low signal-to-noise observations (see e.g. Zafar et al. 2010).

The aim of this paper is to study the dust formation in the nearby bright Type II-P SN 2004dj. This SN occured
in the compact cluster Sandage-96 (S96) in NGC~2403, and {\bf owing to} its proximity ($\sim$ 3.5 Mpc, Vink\'o et al. 2006)
it has been intensively studied since the discovery by K. Itagaki (Nakano et al. 2004, Patat et al. 2004). 
Early observations have been summarized by Vink\'o et al. (2006, hereafter Paper~I), while the physical properties of
S96 have been recently discussed by Vink\'o et al. (2009, hereafter Paper~II). The progenitor was identified as a massive
(12 $M_{\odot} \lesssim M_{prog} \lesssim$ 20 $M_{\odot}$) star within S96 (Ma{\'{\i}}z-Apell{\'a}niz et al.
2004; Wang et al. 2005; Paper~II). 

The evolution of SN~2004dj was extensively observed by {\it Spitzer}. The earliest observations have been 
presented by Kotak et al. (2005). In the following sections we analyze the {\it Spitzer} and 
{\it Hubble} observations that show signs of dust formation (Section~\ref{obs} and Section~\ref{anal}), 
then we compare various dust models with these observations in Section~\ref{model}. 
Finally, we present our conclusions in Section~\ref{sum}.   

\section{Observations and data reduction}\label{obs}

This section contains the description of the {\it Spitzer}- and {\it Hubble} observations
that we use to probe the newly formed dust in SN~2004dj. Throughout this paper 
we adopt JD 2,453,187.0 (2004-06-30) as the moment of explosion (see Paper~I).   

\subsection{MIR Photometry with {\it Spitzer}}\label{obs_phot}

We collected all public archival {\it Spitzer} data on SN~2004dj for 
each channel (3.6, 4.5, 5.8 and 8.0 $\mu$m) of the Infrared Array Camera (IRAC) and for 
the 24 $\mu$m channel of the Multiband Imaging Spectrometer (MIPS). 
We use all available basic- and post-basic
calibrated data (BCD and PBCD) obtained from +98 to +1381 days past explosion
(summarized in Table~\ref{tab:phot}). 

\begin{figure*}
\begin{center}
\includegraphics[width=8cm,height=8cm]{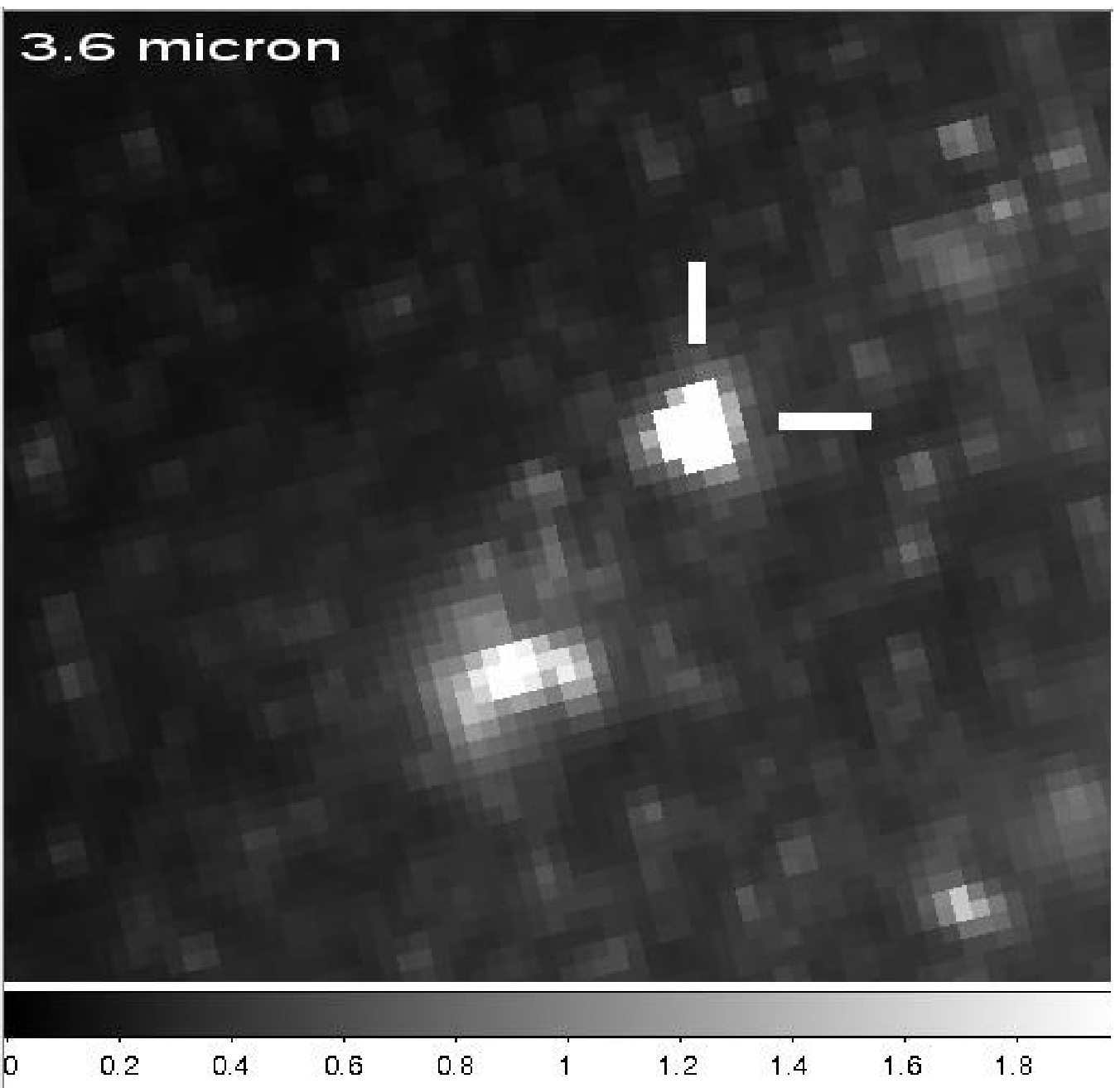} \hskip 3mm
\includegraphics[width=8cm,height=8cm]{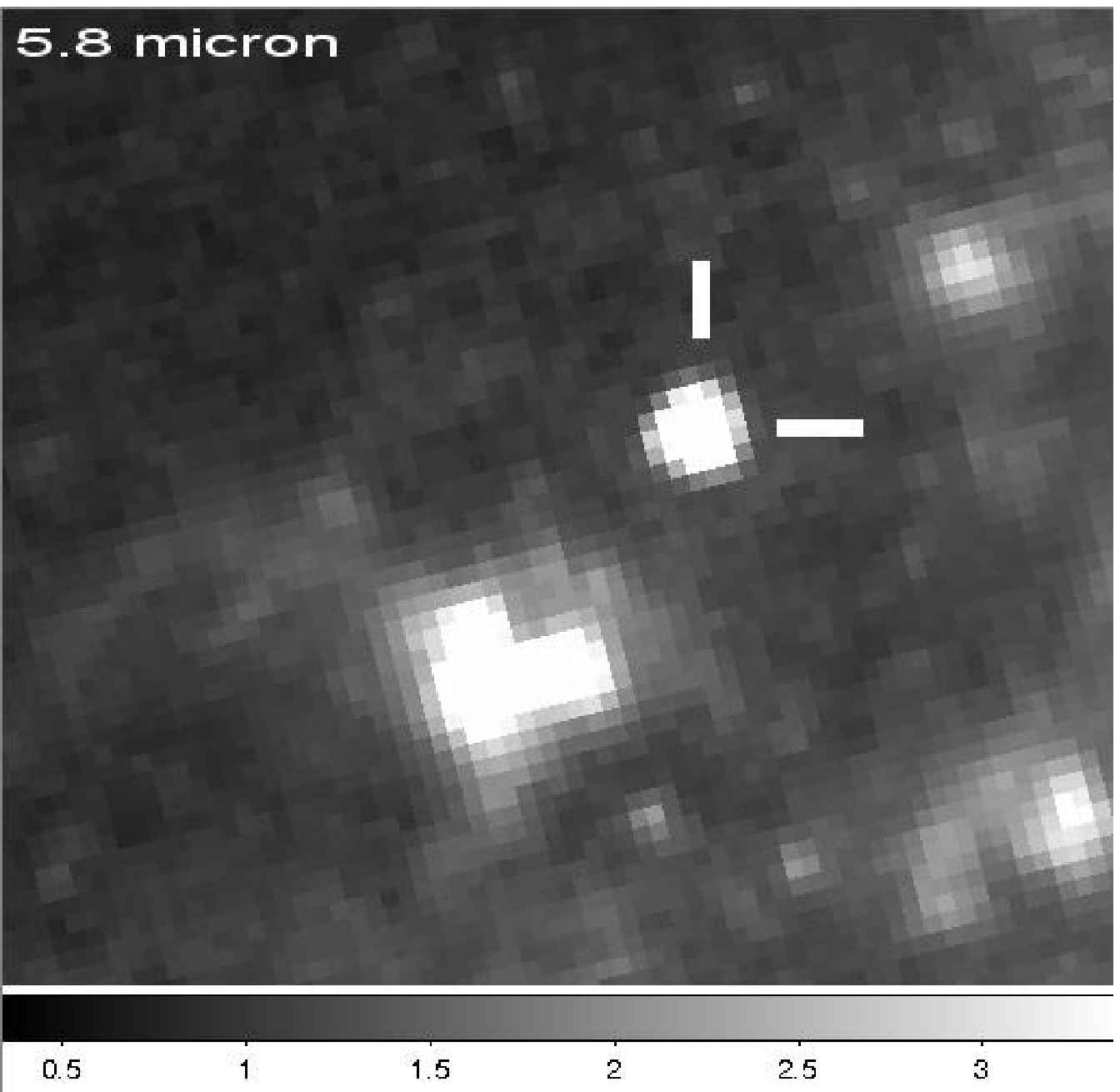}
\vskip 3mm
\includegraphics[width=8cm,height=8cm]{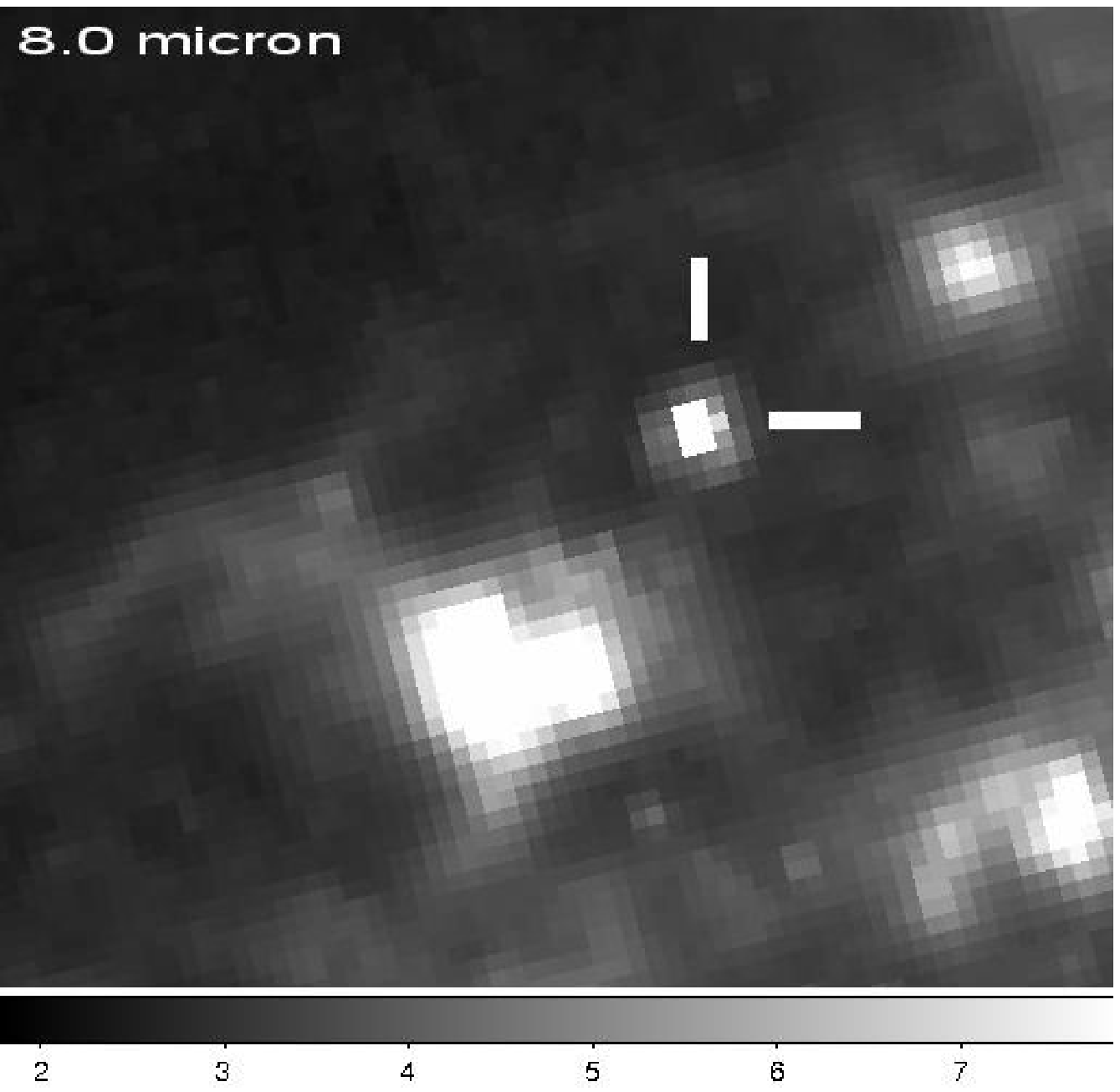} \hskip 3mm
\includegraphics[width=8cm,height=8cm]{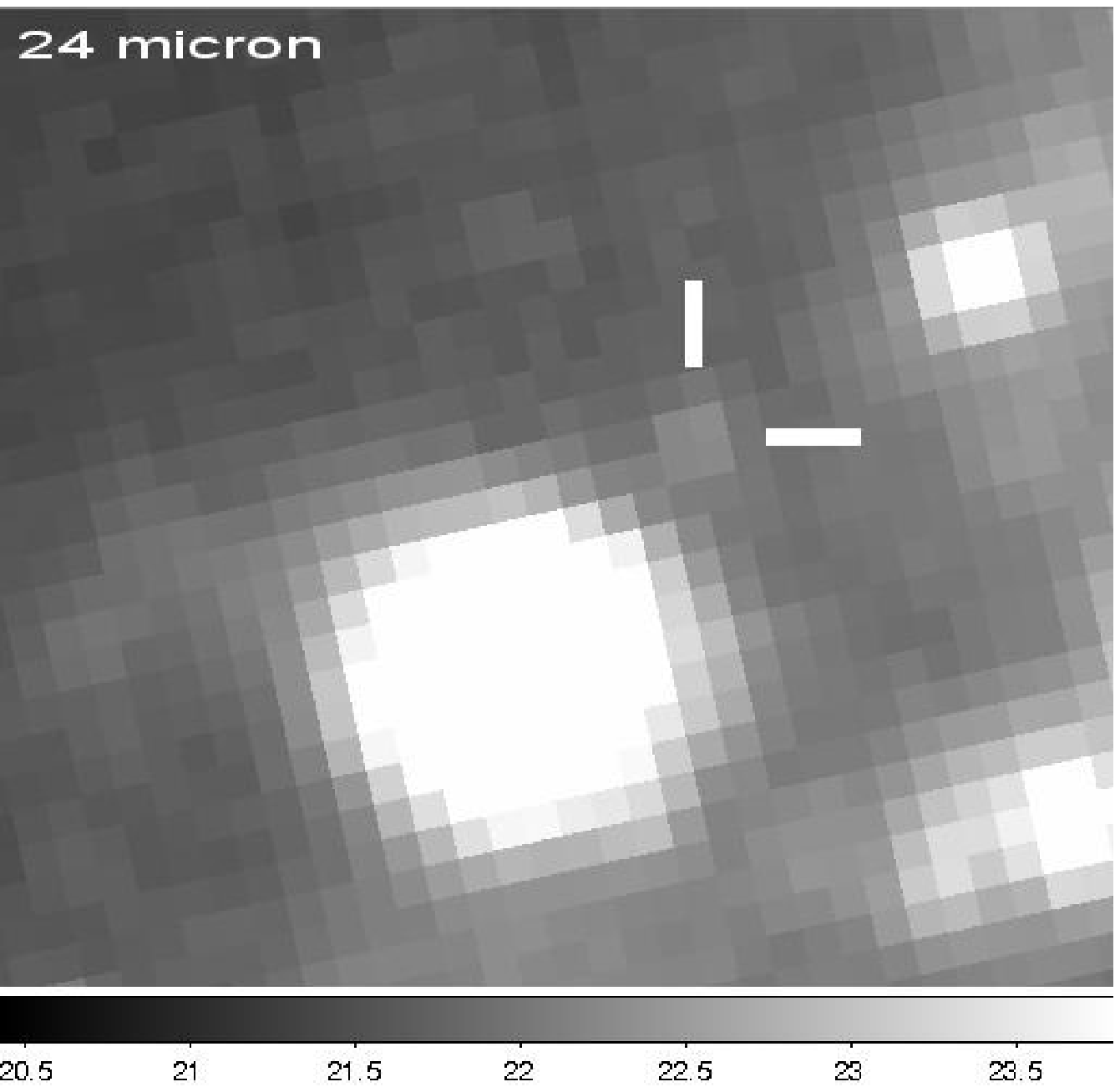}
\end{center}
\caption{Post-BCD images showing the environment of SN 2004dj obtained with {\it Spitzer} on 2004-10-12: 
IRAC 3.6 micron (top left), 5.8 micron (top right), 8.0 micron (bottom left), and MIPS 24.0 micron (bottom
right). The SN is marked in each panel. The target is very faint on the MIPS image, 
which illustrates the necessity of PSF-photometry (see text for details).}
\label{fig:pbcd_img}
\end{figure*}

The IRAC BCD images were processed using the {\it IRACproc} software
(Schuster et al. 2006) to create final mosaics with a scale of
0.86$^{\prime\prime}$/pixel. Aperture photometry was also carried out with this package
(which uses built-in IRAF\footnote{IRAF is distributed by the National Optical Astronomy 
Observatories, which are operated by the Association of Universities for Research in Astronomy, Inc., under cooperative 
agreement with the National Science Foundation.} scripts) with an aperture radius 2$^{\prime\prime}$ and a
background annulus from 2$^{\prime\prime}$ to 6$^{\prime\prime}$. Aperture corrections of 1.213, 1.234,
1.379, and 1.584 were used for channels 1-4, respectively (Table 5.7 in IRAC
Data Handbook, Reach et al. 2006). We compared our results against
aperture photometry on PBCD images using the IRAF {\it phot} task with the same parameters,
and found a reasonably good agreement.

Registration, mosaicing and resampling of the MIPS images were made by
using the DAT software (Engelbracht et al. 2007). The source was not
detected on the 70 $\mu$m images, therefore we analyzed only the 24 $\mu$m frames.
The photometry of the 24 $\mu$m data was complicated, because the source 
was very faint, peaking only slightly above background (see Fig.\ref{fig:pbcd_img}).
First, aperture photometry was performed on the 1.245$^{\prime\prime}$/pixel scale mosaics using
an aperture radius of 3.5$^{\prime\prime}$ and a background annulus from 6$^{\prime\prime}$ to 8$^{\prime\prime}$. An aperture
correction of 2.78 was then applied (Engelbracht et al. 2007). The choice of
such a small aperture and annulus was motivated by the
close, bright \ion{H}{II} region, which makes the background high and variable on 
one side of the source.    

Second, PSF-photometry was carried out on the 24 $\mu$m MIPS frames using a custom-made
empirical PSF (assembled from multiple epoch observations of HD173398)
using {\it IRAF/DAOPHOT}. We set the PSF radius as 90 pixels, for which no computed aperture 
correction is available. Therefore, the photometric 
calibration was done by comparing the results of the properly calibrated 
aperture photometry and our PSF-photometry for two isolated point-like 
sources within the SN-field that do not suffer from the contamination of
a nearby bright object. We compared the the sum of the count rates on the DAT frames (in DN/s)
from both aperture- and PSF-photometry, aperture-corrected the
numbers from aperture photometry and found the following linear relationship
betweeen the aperture-corrected count rates ($f_{apcor}$) 
and those from our PSF-photometry ($f_{PSF}$):

\begin{equation}
f_{apcor} = 1.183(\pm 0.012) \times f_{PSF} + 16.53(\pm 16.91) .
\end{equation}
\noindent

The fluxes from PSF-photometry corrected this way were then converted
to mJy as prescribed for MIPS and found to be within the
errors of those from aperture photometry. 

We also performed PSF-photometry with the {\it IDP3} software (Stobie \& Ferro 2006), which agreed well
with the results of {\it DAOPHOT}. As a triple-check, we
also performed aperture photometry on the PBCD-images and found reasonable agreement
between the absolute flux levels computed from PSF- and aperture photometry. 
Thus, the final 24 $\mu$m fluxes were calculated as the average of the results from the 
different methods. The final photometry of SN~2004dj is collected in Table~\ref{tab:phot}. 
The errors represent the standard deviation of the data from the three different photometry methods.   

Kotak et al. (2005) presented some MIR photometric points observed at
$\sim$100 days after explosion. Their reported flux values are
generally consistent with ours, but there are some differences (mainly in
channels 4.5 and 24 $\mu$m). The reason for these minor differences may be 
that they applied larger aperture radii (3.6" and 5.6" for 4.5 and 24 $\mu$m data,
respectively) and performed only aperture photometry on the PBCD images.

\begin{table*}
\caption{\label{tab:phot} Spitzer photometry for SN~2004dj. $t_{expl}$ = 2,453,187 was adopted as the moment
of explosion.}
\centering
\newcommand\T{\rule{0pt}{3.1ex}}
\newcommand\B{\rule[-1.7ex]{0pt}{0pt}}
\begin{tabular}{lcccccccc}
\hline
\hline
~ &~ &~ & \multicolumn{6}{c}{Flux (10$^{-20}$ erg s$^{-1}$ cm$^{-2}$ \AA$^{-1}$)} \T \B \\
\cline{4-9}
UT Date & MJD $-$ & $t - t_{expl}$ & \multicolumn{4}{c}{IRAC} & IRS PUI & MIPS \T \\
~ & 2450000 & (days) & 3.6 $\mu$m & 4.5 $\mu$m & 5.8 $\mu$m & 8.0 $\mu$m & 13.0 - 18.5 $\mu$m & 24 $\mu$m \B \\
\hline
2004-10-07\tablefootmark{a} & 3285.6 & 98 & 24800(59) & 13900(51) & 5520(53) & 1940(16) & ... & ... \T \\
2004-10-08\tablefootmark{b} & 3286.9 & 99 & 24000(59) & 14000(32) & 5580(29) & 1830(16) & ... & ... \\
2004-10-12$^{b}$ & 3290.6 & 103 & 18300(61) & 13600(30) & 5000(29) & 1530(15) & ... & ... \\
2004-10-12$^{b}$ & 3291.4 & 104 & ... & ... & ... & ... & ... & 45(3) \\
2004-10-14$^{a}$ & 3293.0 & 106 & ... & ... & ... & ... & ... & 50(6) \\
2004-10-16$^{b}$ & 3295.0 & 108 & ... & ... & ... & ... & ... & 31(2) \\
2004-11-01$^{a}$ & 3310.6 & 123 & 9680(39) & 10700(32) & 3960(25) & 1090(11) & ... & ... \\
2004-11-06$^{a}$ & 3316.2 & 129 & ... & ... & ... & ... & ... & 55(2) \\
2005-03-03$^{a}$ & 3432.8 & 245 & ... & ... & ... & ... & ... & 48(2) \\
2005-03-24$^{a}$ & 3454.4 & 267 & 2740(22) & 4910(17) & 1280(18) & 551(11) & ... & ... \\
2005-04-01$^{a}$ & 3462.5 & 275 & ... & ... & ... & ... & ... & 44(2) \\
2005-10-20\tablefootmark{c} & 3664.2 & 477 & 4110(22) & 3280(12) & 1910(21) & 1030(13) & ... & ... \\
2005-11-22$^{c}$ & 3696.9 & 510 & ... & ... & ... & ... & 138(1) & ... \\
2006-03-23$^{c}$ & 3818.4 & 631 & 2980(20) & 2430(11) & 1960(16) & 1110(12) & ... & ... \\
2006-04-23$^{c}$ & 3848.4 & 662 & ... & ... & ... & ... & 182(1) & ... \\
2006-10-28\tablefootmark{d} & 4036.7 & 849 & 1870(18) & 1610(9) & 1490(13) & 919(12) & ... & ... \\
2006-10-31\tablefootmark{e} & 4039.6 & 852 & 1850(53) & 1580(39) & 1310(63) & 945(28) & ... & ... \\
2006-11-16$^{d}$ & 4054.7 & 867 & ... & ... & ... & ... & 173(4) & ... \\
2006-11-16$^{e}$ & 4055.0 & 868 & ... & ... & ... & ... & 172(2) & ... \\
2006-12-01$^{e}$ & 4070.8 & 883 & ... & ... & ... & ... & ... & 66(4) \\
2007-04-02$^{e}$ & 4193.3 & 1006 & 1070(16) & 917(10) & 960(14) & 650(11) & ... & ... \\
2007-04-02$^{d}$ & 4193.3 & 1006 & 1080(16) & 922(11) & 870(15) & 602(10) & ... & ... \\
2007-04-13$^{e}$ & 4203.6 & 1016 & ... & ... & ... & ... & ... & 53(4) \\
2007-10-24\tablefootmark{f} & 4397.8 & 1210 & ... & ... & ... & ... & ... & 59(4) \\
2007-10-24\tablefootmark{g} & 4397.8 & 1210 & ... & ... & ... & ... & ... & 58(6) \\
2007-11-04$^{f}$ & 4408.3 & 1221 & ... & ... & ... & ... & 117(3) & ... \\
2007-11-19$^{e}$ & 4423.8 & 1236 & 779(15) & 616(8) & 619(11) & 437(10) & ... & ... \\
2007-11-23$^{f}$ & 4427.6 & 1240 & 771(15) & 613(8) & 619(11) & 444(10) & ... & ... \\
2007-11-24$^{g}$ & 4428.5 & 1241 & 787(14) & 617(7) & 647(7) & 444(10) & ... & ... \\
2007-11-29$^{e}$ & 4433.9 & 1246 & ... & ... & ... & ... & ... & 62(3) \\
2007-12-15$^{g}$ & 4449.9 & 1262 & ... & ... & ... & ... & 113(2) & ... \\
2008-04-07$^{g}$ & 4564.3 & 1377 & 723(15) & 520(6) & 544(11) & 399(11) & ... & ... \\
2008-04-12$^{f}$ & 4568.7 & 1381 & 718(16) & 526(7) & 527(13) & 413(11) & ... & ... \B \\
\hline
\end{tabular}
\tablefoot{\\
\tablefoottext{a}{PID 226 Van Dyk et al. (SONS)} \\
\tablefoottext{b}{PID 159 Kennicutt et al. (SINGS); 24.0 $\mu$m values are MIPSScan data} \\
\tablefoottext{c}{PID 20256 Meikle et al. (MISC)} \\
\tablefoottext{d}{PID 30292 Meikle et al. (MISC)} \\
\tablefoottext{e}{PID 30494 Sugerman et al. (BEKS)} \\
\tablefoottext{f}{PID 40010 Meixner et al. (SEEDS)} \\
\tablefoottext{g}{PID 40619 Kotak et al. (MISC)}
}
\end{table*}

\subsection{MIR Spectroscopy with {\it IRS}}\label{obs_spec}

SN~2004dj was observed with the Infrared Spectrograph
(IRS) onboard {\it Spitzer} on seven epochs between 
October 2004 and November 2006, from +115 to +868 days after
explosion. These observations were made in the
IRSStare mode using the Short-Low (SL) setup.
Table~\ref{tab:spec} contains the
list of spectroscopic data downloaded from the
{\it Spitzer} archive. In addition, several imaging
observations were made with the blue array (16 $\mu$m central wavelength) 
of the peak-up imaging (PUI) mode of IRS, which are listed 
in Table~\ref{tab:phot}.
 
We analyzed the PBCD-frames containing
the spectroscopic data using the SPitzer IRS Custom
Extraction software 
({\it SPICE}\footnote{http://ssc.spitzer.caltech.edu/dataanalysistools/tools/spice/}).
Sky subtraction and bad pixel removal were performed
using two exposures containing the spectrum at different locations, and
subtracting them from each other. Order extraction, wavelength- and
flux-calibration were performed applying built-in templates 
within {\it SPICE}. Finally, the spectra from the 1st, 2nd and
3rd orders were combined into a single spectrum with the overlapping 
edges averaged. In a few cases the sky near the order edges 
was oversubtracted because of excess flux close to the source, 
which resulted in spurious negative flux values in the extracted spectrum. These 
were filtered out using the fluxes in the same wavelength region that were extracted 
from the adjacent orders. The results were also checked by comparing
the extracted fluxes with photometry (Table~\ref{tab:phot}). Reasonable
agreement between the spectral and photometric fluxes was found for 
all spectra.   

\begin{figure}
\begin{center}
\resizebox{\hsize}{!}{\includegraphics{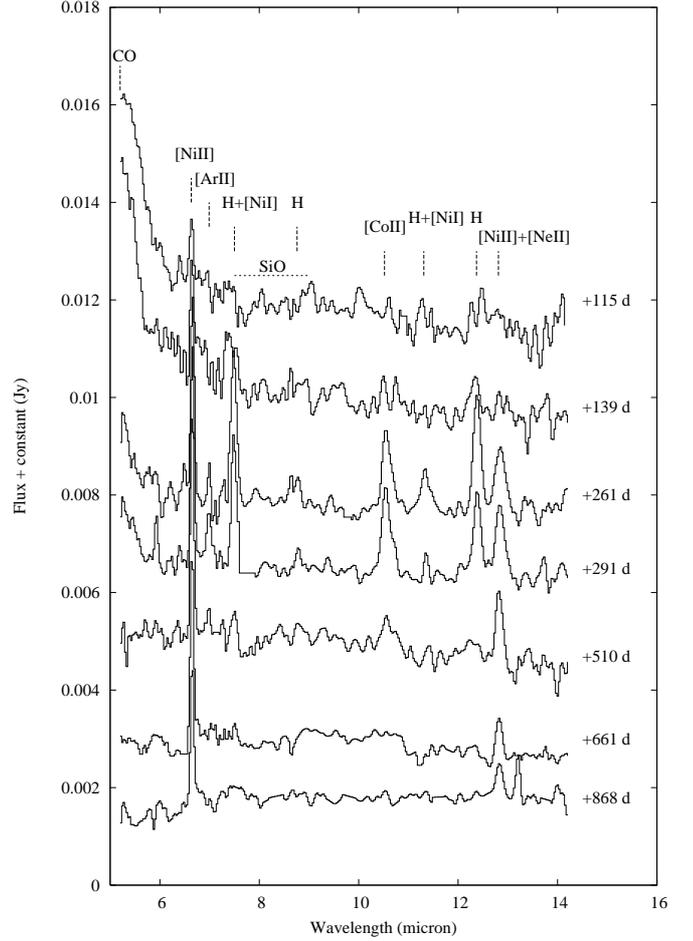}}
\caption{IRS spectra of SN 2004dj in the nebular phase. Line identification is based on 
Kotak et al. (2005, 2006). The evolution of the features is discussed in Sect. \ref{specanal}.}
\label{fig:irs}
\end{center}
\end{figure}

The final calibrated and combined spectra cover the 5.15 - 14.23 $\mu$m 
wavelength regime with resolving power $R$ $\sim 100$. 
These are plotted in Fig.\ref{fig:irs}, where a small vertical
shift is applied between the spectra  for better visibility. 
The analysis of the spectral features and evolution will be presented
in Section~\ref{anal}. 

The fluxes of SN~2004dj in the PUI frames were measured
via aperture photometry by the MOsaicker and Point source EXtractor 
({\it MOPEX}\footnote{http://ssc.spitzer.caltech.edu/dataanalysistools/tools/mopex/})    
software. The results are given in Table~\ref{tab:phot}. These
broad-band integrated fluxes were only used for checking the
shape of the SED between the IRAC and MIPS bands, but were not
included in the detailed analysis in Section~\ref{anal}. 

\begin{table*}
\caption{\label{tab:spec} IRS observations of SN~2004dj}
\centering
\newcommand\T{\rule{0pt}{3.1ex}}
\newcommand\B{\rule[-1.7ex]{0pt}{0pt}}
\begin{tabular}{lcccl}
\hline
\hline
UT Date & MJD $-$ & $t-t_{expl}$ & ID & Proposal ID (PID) \T \\
  & $2,450,000$ & (days) & & \B \\
\hline
2004-10-24 & 3302.3 & 115 & r12113152 & 226 (SONS, Van Dyk) \T \\
2004-11-16 & 3325.8 & 139 & r12114432 & 226 (SONS, Van Dyk) \\
2005-03-18 & 3447.6 & 261 & r12113408 & 226 (SONS, Van Dyk) \\
2005-04-17 & 3477.9 & 291 & r12115456 & 226 (SONS, Van Dyk) \\
2005-11-22 & 3696.8 & 510 & r14458880 & 20256 (MISC, Meikle) \\
2006-04-23 & 3848.4 & 661 & r14465280 & 20256 (MISC, Meikle) \\
2006-11-16 & 4055.1 & 868 & r17969152 & 30292 (MISC, Meikle) \B \\
\hline
\end{tabular}
\end{table*}

\begin{table*}
\caption{\label{tab:speclines} Identified lines in the IRS spectra of SN~2004dj. "+" signs represent
positive detection, while "-" symbols mean non-detection of the given feature at the
specified epochs.}
\centering
\newcommand\T{\rule{0pt}{3.1ex}}
\newcommand\B{\rule[-1.7ex]{0pt}{0pt}}
\begin{tabular}{llccccccc}
\hline
\hline
$\lambda$ ($\mu$m) & Ion & +115d & +139d & +261d & +291d & +510d & +661d & +868d \T \B \\
\hline
4.55   & CO & + & + & + & + & - & - & - \T \\
6.6342 & [\ion{Ni}{II}] & + & + & + & + & + & + & + \\
6.9852 & [\ion{Ar}{II}] & - & - & + & + & + & - & - \\
7.4578 & H-Pf$\alpha$& + & + & + & - & - & - & - \\
7.5005 & H-Hu$\beta$& + & + & + & + & + & + & - \\
7.5066 & [\ion{Ni}{I}] & \multicolumn{7}{c}{blend with H-Hu$\beta$} \\
8.7577 & H 7-10 & + & + & + & + & - & - & - \\
10.521 & [\ion{Co}{II}] & + & + & + & + & + & - & - \\
11.306 & H 7-9 & \multicolumn{7}{c}{blend with [\ion{Ni}{I}]} \\
11.308 & [\ion{Ni}{I}] & + & - & + & + & - & - & - \\
12.368 & H-Hu$\alpha$ & + & + & + & + & + & - & - \\
12.729 & [\ion{Ni}{II}] & \multicolumn{7}{c}{blend with [\ion{Ne}{II}]} \\
12.813 & [\ion{Ne}{II}] & + & + & + & + & + & + & + \B \\
\hline
\end{tabular}
\end{table*}

\subsection{Imaging polarimetry with {\it HST}}\label{obs_polar}

SN~2004dj and its surrounding cluster, Sandage-96, were imaged by {\it HST}/ACS High-Resolution
Camera (HRC) on 2005 Aug 28 
(proposal GO-10607, PI: B.E.K. Sugerman), +425 days after explosion. Among others, 
three sets of four drizzled frames were recorded through the $F435W$ filter at 0, 60, and 
120 degree position of the $POLUV$ polarization filter 
(see Paper~II for the description and analysis of all observations). Here we consider
only the polarization measurement of the SN. 

We measured the flux from SN~2004dj with PSF-photometry using the {\it DOLPHOT} software
(Dolphin 2000) the same way as described in Paper~II. From the fluxes measured at
three different polarizer angles ($I_0$, $I_{60}$ and $I_{120}$), the Stokes-parameters
($I$, $Q$, $U$) were calculated applying the formulae given by e.g. Sparks \& Axon (1999). 
The degree-of-polarization, $p$ was then derived as
\begin{equation}
p = \frac{\sqrt{Q^2 + U^2}}{I}.
\end{equation}

This resulted in a measured degree-of-polarization $p$ = $0.0941 \pm 0.0029$, which is, of course,
much higher than the true $p$ from SN~2004dj, owing to instrumental polarization 
($p_i$), and the polarization from interstellar matter ($p_{ISM}$).  

The instrumental polarization of ACS was studied by Biretta et al. (2004). 
Using three sets of in-orbit calibration data they measured the
instrumental polarization as $p_{i} = 0.086 \pm 0.002$ for the 
ACS/HRC+$F435W$+$POLUV$ detector+filter combination (see their Table~20). 
They also give a semi-empirical formula describing the relative uncertainty
of the degree-of-polarization, $\sigma_p / p$, as a function of the 
signal-to-noise of the flux measurement.

Adopting this calibration, the measured true degree-of-polarization, corrected
for the instrumental zero-point, is $p_{true} = 0.0081 \pm 0.0036$. 
The average S/N of the three flux measurements for SN~2004dj is $\sim 200$,
from which we derived $\sigma_p / p = 0.5588$, thus, $\sigma_p = 0.0045$, which 
agrees well with the error estimate of $p_{true}$ given above. 

Further analysis and discussion of the polarization data are presented in 
Section~\ref{anal}.

\section{Analysis of the observations}\label{anal}

\subsection{The effect of the surrounding cluster Sandage-96}\label{anal_s96}

The bright, compact cluster S96 surrounding SN~2004dj (Paper II) made late-time
optical measurements difficult, because the light from the cluster stars contribute significantly
to the total measured fluxes. This was very problematic in the optical, but fortunately
less severe in the MIR, where the cluster is much fainter.
Although we have no pre-SN MIR photometry for S96, we estimated 
its MIR flux using the best-fitting cluster models from Paper~II. 
These models are simple, coeval stellar populations (SSP), but we considered
three different model sets using different input physics to test the model dependence
of the flux predictions. 

\begin{figure}
\begin{center}
\resizebox{\hsize}{!}{\includegraphics{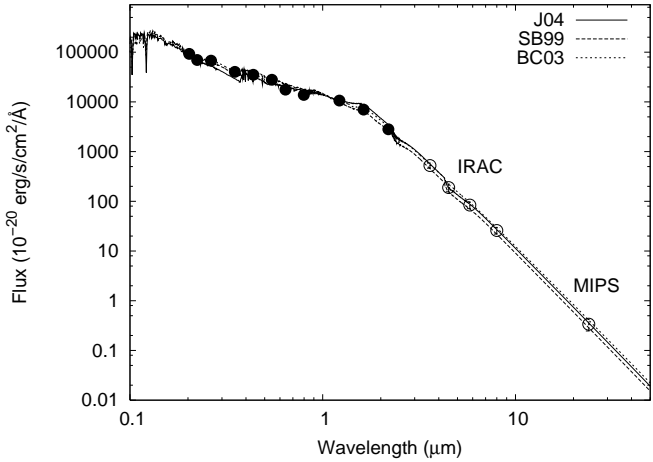}}
\caption{Best-fitting model SEDs for the host cluster Sandage-96 from Paper~II. Filled circles
represent observed data for the cluster, while open symbols denote the averaged MIR-fluxes
from the models (see Table~\ref{tab:s96con}).}
\label{fig:s96sed}
\end{center}
\end{figure}

Figure \ref{fig:s96sed} illustrates the MIR SEDs for the three SSP models that 
produced the best fit to the observed UV+optical+NIR data of S96 
(Fig.\ref{fig:s96sed} filled symbols; see Paper~II for details).    
Fortunately, the three models predicted very similar MIR fluxes in the 
{\it Spitzer} bands. Their average values (in cgs units) are collected in Table~\ref{tab:s96con},
where the errors illustrate the flux differences between the models. 

\begin{table}
\begin{center}
\caption{\label{tab:s96con} Contribution of the Sandage-96 cluster to the MIR fluxes of SN 2004dj}
\newcommand\T{\rule{0pt}{3.1ex}}
\newcommand\B{\rule[-1.7ex]{0pt}{0pt}}
\begin{tabular}{c|cc}
\hline
\hline
Wavelength & Model flux & Flux error \T \\
($\mu$m) & \multicolumn{2}{c}{(erg s$^{-1}$ cm$^{-2}$ \AA$^{-1}$)} \B \\
\hline
IRAC 3.6 & $5.25 \times 10^{-18}$ & $6.2 \times 10^{-19}$\T \\
IRAC 4.5 & $1.89 \times 10^{-18}$ & $3.4 \times 10^{-19}$ \\
IRAC 5.8 & $8.40 \times 10^{-19}$ & $1.3 \times 10^{-19}$ \\
IRAC 8.0 & $2.59 \times 10^{-19}$ & $0.4 \times 10^{-19}$ \\
MIPS 24.0 & $3.37 \times 10^{-21}$ & $0.9 \times 10^{-21}$\B \\
\hline
\end{tabular}
\end{center}
\end{table}

Evidently the cluster contributes significantly to the measured SN fluxes 
only in the IRAC channels,
but the cluster flux drops below the observational uncertainties longward of 10 $\mu$m. 
Thus, we subsequently corrected the IRAC fluxes for this offset using the
model fluxes from Table~\ref{tab:s96con}, but this correction was not necessary
for the PUI and MIPS data.  

\subsection{Spectral features and evolution}\label{specanal}

The nebular spectra of SN~2004dj presented in Fig.\ref{fig:irs} are typical for Type II-P SNe. They show features and evolution 
very similar to the available MIR spectra of a few SNe, including SN~1987A (Wooden et al. 1993; 
Roche et al. 1993), SN~2005af (Kotak et al. 2006) and SN~2004et (Kotak et al. 2009). 
The first two spectra of SN~2004dj (taken at +115 d and +139 d) have been already 
presented and discussed by Kotak el al. (2005). We include these spectra in the present study 
for completeness, but the conclusions for them are generally the same as those of Kotak et al. (2005). 

The spectra consist of permitted emission lines of \ion{H}{I} (Pfund-, Humphrey- and n=7 series), and
forbidden lines of [\ion{Ni}{I}], [\ion{Ni}{II}], [\ion{Co}{II}] and [\ion{Ar}{II}]. Table~\ref{tab:speclines}
summarizes the identified lines, based on Wooden et al. (1993) and Kotak et al. (2005). 
The presence of [\ion{Ne}{II}], forming a blend with [\ion{Ni}{II}] at 12.8 $\mu$m, 
was suggested by Kotak et al. (2006) for SN~2005af. This feature is also
present in SN~2004dj for all phases covered by {\it IRS}-observations, although the two components
cannot be resolved. 

The blue edge of the first two spectra is influenced by the red wing of the 1-0 vibrational transition of 
the CO-molecule, similar to SN~1987A, as first pointed out by Kotak et al. (2005). CO 1-0 remained detectable in
the two subsequent spectra up to +291 days. No sign of the SiO fundamental band
around 8 $\mu$m was detected in SN~2004dj, unlike in SNe 1987A, 2005af and 2004et. Because we present
evidence that SN~2004dj showed significant dust formation after +400 days, 
the absence of detectable SiO can be an important constraint for the chemical composition of the
newly formed dust grains.       

\begin{table}
\caption{\label{tab:ni} Measured line fluxes and ionization fraction ($x$) for $^{58}$Ni and $^{56}$Co}
\centering
\begin{tabular}{lcccc}
\hline
\hline
Phase & [\ion{Co}{II}]\tablefootmark{a} & [\ion{Ni}{II}] & [\ion{Ni}{I}]\tablefootmark{b} & $x$($^{58}$Ni)\\
(days) & 10.52 $\mu$m & 6.63 $\mu$m & 7.51 $\mu$m & \\
\hline
115 & 1.6 & 8.47 & -- & --\\
139 & -- & 17.8 & -- & --\\
261 & 16.1 & 28.4 & 2.5 & 0.87 \\
291 & 16.8 & 27.5 & 2.0 & 0.89 \\
510 & 8.3 & 27.9 & 1.86 & 0.89 \\
661 & -- & 25.4 & 1.88 & 0.88 \\
868 & 1.1 & 16.7 & -- & -- \\
\hline
\end{tabular}
\tablefoot{\\
\tablefoottext{a}{$\times 10^{-15}$ erg s$^{-1}$ cm$^{-2}$} \\
\tablefoottext{b}{corrected for Pa$\alpha$, Hu$\beta$ and \ion{H}{I} 7-11 contribution (see text)}
}
\end{table}

\subsection{Masses of Ni and Co}\label{nico}

Masses of freshly synthesized Ni estimated from observations may provide important constraints 
for SN explosion models (e.g. Woosley \& Weaver 1995; Thielemann et al. 1996; Chieffi \& Limongi 2004;
Nomoto et al. 2006).
Following the methods presented in Roche et al. (1993) and Wooden et al. (1993), we derived 
the ionization fraction of stable $^{58}$Ni from the measured ratio of fluxes of the collisionally-excited lines of
[\ion{Ni}{II}] at 6.63 $\mu$m and [\ion{Ni}{I}] at 7.51 $\mu$m. It was shown by Wooden et al. (1993) and Kotak et al.
(2005) that the critical density (at which the collisional de-excitation rate equals the rate of spontaneous
decay) for the 6.63 $\mu$m line is $\sim 1.3 \times 10^{7}$ cm$^{-3}$, while the electron density $\sim 1$
year after explosion is $\sim 10^{8}$ cm$^{-3}$, one order of magnitude higher, making LTE-conditions valid. 
Under these conditions the total number of excited atoms at the upper level $u$ of the given transition can
be expressed as
\begin{equation}
N_u ~=~ 4 \pi D^2 F_{ul} \lambda / h c A_{ul},
\end{equation}
where $D$ is the distance to the SN (3.5 Mpc, Paper~I), $F_{ul}$ is the measured line flux 
(in erg s$^{-1}$ cm$^{-2}$), $\lambda$ is the wavelength of the transition and $A_{ul}$
is the transition probability (in cm$^{-1}$). Note that this expression is valid for optically
thin lines, but the [\ion{Ni}{II}] 6.63 $\mu$m feature was probably optically thick between 
100-660 days (Wooden et al., 1993). Thus, the derived quantities for this line 
are actually lower limits. The ratio of the total number of neutral Ni$^0$ and ionized Ni$^+$ 
can be obtained from the ratio of measured line fluxes via Boltzmann- and Saha-equations.  

It is important that the 7.51 $\mu$m line in the {\it IRS} spectra is an unresolved blend
of [\ion{Ni}{I}] with Pa$\alpha$, Hu$\beta$ and \ion{H}{I} 7-11, which makes the direct measurement of the [\ion{Ni}{I}] 
feature difficult. This was already noted by Kotak et al. (2005), but in a subsequent paper Kotak
et al. (2006) measured the flux of this line directly and derived the ionization fraction 
$x$ = Ni$^+$ / (Ni$^0$ + Ni$^+$) $~\sim0.4$ for SN~2004dj at +261 days. This is roughly a factor of 2 lower than 
what was measured in SN~1987A at that epoch ($x \sim 0.8$, Wooden et al. 1993). In order to 
clarify this, we estimated the line flux of the [\ion{Ni}{I}] feature assuming that the \ion{H}{I} 
features have the same ratio to the total line flux as was measured by Wooden et al. (1993) 
for SN~1987A in higher resolution spectra. The contribution from \ion{H}{I} was found to be as high as
90 percent at +260 days, which decreased to 75 percent at +450 days and less than 10 percent
at +660 days. We used these numbers to estimate the [\ion{Ni}{I}] fluxes from the measured
total line fluxes (the [\ion{Ni}{II}] 6.63 $\mu$m feature was not affected by this complication). 
The results are collected in Table~\ref{tab:ni}. Using these line fluxes and assuming 
$T_e = 3000$ K electron temperature (Wooden et al., 1993), the ionization fraction was computed as
\begin{eqnarray}
y = {N({\rm Ni^+}) \over N({\rm Ni^0})} = \nonumber \\
{F(6.63) \over F(7.51)} {6.63 \over 7.51} { {g_{7.51} A_{7.51}} \over {g_{6.63} A_{6.63}}} {z^{+}(T_e) \over z^{0}(T_e)} \exp[(E_{6.63}-E_{7.51})/kT_e],
\end{eqnarray}
then applying $x ~=~ y/(1+y)$, where $g$ is the statistical weight, $z^{+,0}(T)$ are the partition functions
for ionized and neutral Ni, respectively, and $E$ is the excitation potential for the upper level of the given transition. 
The values of the atomic parameters were adopted from 
the NIST Atomic Spectral Database\footnote{http://www.nist.gov/physlab/data/asd.cfm}. 
The derived ionization fractions are shown in the last column of Table~\ref{tab:ni}. These values
are very close to those obtained by Wooden et al. (1993) for SN~1987A, but significantly higher
than the results by Kotak et al. (2006). It is probable that Kotak et al. overestimated
the contribution of [\ion{Ni}{I}] to the 7.51 $\mu$m feature, and therefore underestimated the ionization 
fraction for Ni. Our new result suggests that the Ni ionization and probably other physical conditions
in the ejecta of SN~2004dj were quite similar to those in SN~1987A.

\begin{figure}
\begin{center}
\resizebox{\hsize}{!}{\includegraphics{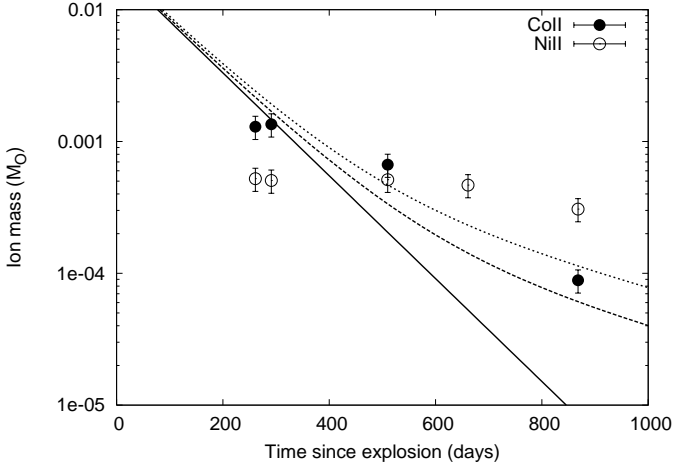}}
\caption{Ni and Co masses in $M_\odot$ calculated from the strength of forbidden emission lines via Eq.3.
Open symbols: Ni-masses from the [\ion{Ni}{II}] 6.63$\mu$ feature; filled symbols: Co-masses
from the [\ion{Co}{II}] 10.52$\mu$ feature. Lines show the expected amount of Co predicted from the
radioactive decay of 0.02 $M_\odot$ radioactive $^{56}$Ni synthesized in the explosion. Continuous line: 
only $^{56}$Co; dashed line: $^{56}$Co plus $^{57}$Co with solar abundance ratio; dotted line:
$^{56}$Co plus $^{57}$Co assuming twice solar abundance ratio.}
\label{fig:nico}
\end{center}
\end{figure}

We also derived the total mass of $^{58}$Ni by applying Eq.3 to the 6.63 $\mu$m feature and assuming
LTE. This resulted in $M$($^{58}$Ni) $\sim 5 \times 10^{-4}$ $M_\odot$, a factor of 2 higher than 
Kotak et al. (2005) obtained from the first two spectra. Note that the mass of the radioactive $^{56}$Ni 
synthesized during the explosion was about $2 \times 10^{-2}$ $M_\odot$ (Paper~I), 
about two orders of magnitude higher (see Fig.\ref{fig:nico}).        

A similar analysis could not be completed for Co, because neither the neutral 
[\ion{Co}{I}] forbidden lines between 3.0 and 3.75 $\mu$m (Wooden et al. 1993), nor
the [\ion{Co}{I}] 12.25$\mu$ feature (Roche et al. 1993) were detectable in the low-resolution 
{\it IRS} spectra. Nevertheless, we derived masses of singly ionized Co via the strength of the 
[\ion{Co}{II}] 10.52$\mu$ transition. This feature was weak in the +115 and +139 day 
spectra, but became one of the
strongest features in the +261 and +291 day spectra. At +510 days it decreased considerably,
and was no longer detectable after +661 days. This behavior is fully consistent with the 
expected radioactive decay of $^{56}$Co. In Fig.\ref{fig:nico} we plot the calculated
Co-masses (filled symbols) together with the predicted masses from the decay
of 0.02 $M_\odot$ radioactive $^{56}$Ni synthesized in the explosion (Paper~I).
Since the observed Co-masses were derived from a single \ion{Co}{II} transition only,
these values are actually lower limits of the total Co-mass.  
Because $^{57}$Co (produced by the decay of $^{57}$Ni) may also contribute to the 
observed feature, its presence was also included in the radioactive model, assuming solar and twice 
solar abundance ratio (dashed and dotted lines, respectively; Roche et al. 1993). 
We conclude that the Co-masses derived from the single [\ion{Co}{II}] 10.52$\mu$ 
transition agree very well with the model assuming that 0.02 $M_\odot$ radioactive
$^{56}$Ni were synthesized in the explosion.

\begin{figure}
\begin{center}
\resizebox{\hsize}{!}{\includegraphics{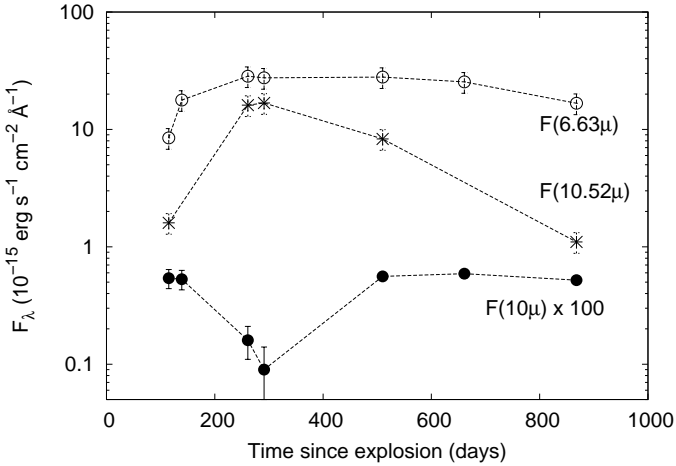}}
\caption{Temporal evolution of the 10 $\mu$m continuum flux, the [\ion{Ni}{II}] 6.63 $\mu$m and
the [\ion{Co}{II}] 10.52 $\mu$m line fluxes.}
\label{fig:irs2}
\end{center}
\end{figure}

\subsection{The continuum emission at 10 $\mu$m}\label{10mu}

In Fig.\ref{fig:irs2} the evolution of the continuum flux at 10 $\mu$m is plotted with
the [\ion{Co}{II}] 10.52$\mu$ and [\ion{Ni}{II}] 6.63$\mu$ line fluxes. There is a striking
similarity between these curves and those of SN~1987A (cf. Fig.2 in Roche et al. 1993). 
The evolution of the [\ion{Ni}{II}] and [\ion{Co}{II}] line fluxes were discussed above. 
The 10 $\mu$m flux values show a significant decline between +115 and 291 days, 
but after that they brighten up and by +510 days they reach the same level as in the earliest spectrum and
only slightly decrease later.

Roche et al. (1993) explained the 10 $\mu$m light curve of SN~1987A as caused by optically thin free-free  
emission by $\sim$300 days, then after $\sim$500 days the increasing flux was interpreted as the sign 
of dust formation. We invoke the same mechanism for SN~2004dj to explain the 10 $\mu$m light curve here.
Similar temporal evolution could be identified in the IRAC and MIPS light curves as well, 
which are discussed below. 


\subsection{Mid-IR light curves}\label{anal_lightc}

\begin{figure}
\begin{center}
\resizebox{\hsize}{!}{\includegraphics{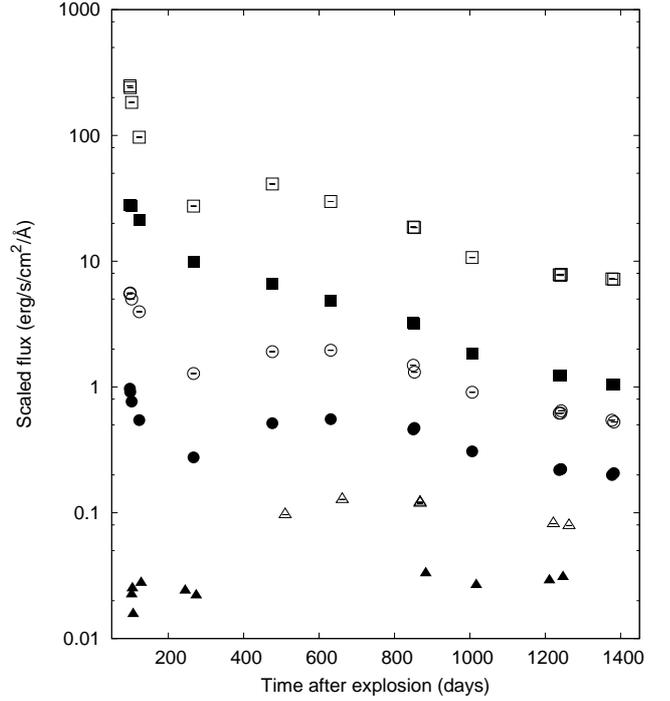}}
\end{center}
\caption{IRAC (3.6 $\mu$m - open squares, 4.5 $\mu$m - filled squares, 5.8 $\mu$m - open circles, 8.0 $\mu$m - filled circles), IRS PUI (open triangles) and MIPS 24.0 $\mu$m (filled triangles) light curves of SN 2004dj. The excess emission appearing on 3.6, 5.8 and 8.0 $\mu$m peaks later at longer wavelengths, which is consistent with a warm, cooling dust formed one year after the core collapse.}
\label{fig:mirlc}
\end{figure}

\begin{figure}
\begin{center}
\leavevmode
\resizebox{\hsize}{!}{\includegraphics{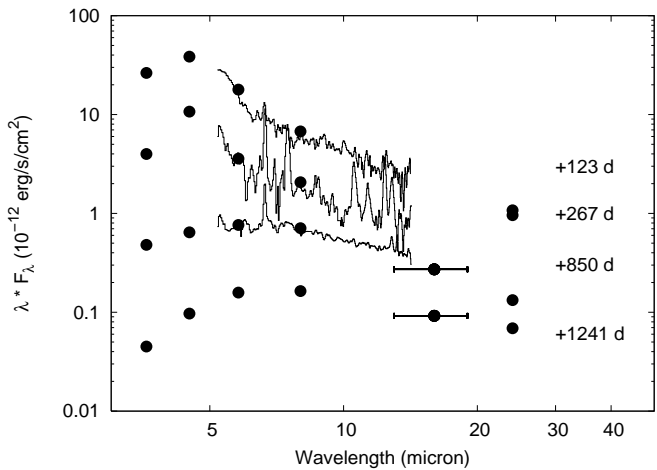}}
\caption{Evolution of MIR SEDs of SN 2004dj}
\label{fig:sedcgs2}
\end{center}
\end{figure}

Similar to the temporal evolution of the 10 $\mu$m continuum flux (Fig.\ref{fig:irs2}), the light curves
of IRAC data from the 3.6 $\mu$m, 5.8 $\mu$m and 8.0 $\mu$m channels also show a significant peak 
after $+400$ days post-explosion (Fig.\ref{fig:mirlc}). Such late-time MIR excess is usually 
considered as a strong evidence for the presence of dust.  
The excess emission peaks later at longer wavelengths, which is 
consistent with a model of warm, cooling dust grains formed in the ejecta. 
Unfortunately, there are no MIPS data during the peak of the IRAC light curves, but the 24 $\mu$m
fluxes also show a slight excess after +800 days with respect to those obtained between +100 - 300 days. 

The 4.5 $\mu$m fluxes do not show the peak that the other IRAC channels do. 
The most plausible explanation for this is the presence of the 1-0 vibrational band of CO at 
4.65 $\mu$m (Section \ref{specanal}) that contributes significantly to the measured flux in the
4.5 $\mu$m channel. After $\sim$ +500 days when the CO band disappeared (Fig.\ref{fig:irs}), the light curve
in this channel behaves similarly to those in the other IRAC bands. 

In Fig.\ref{fig:sedcgs2} the evolution of the MIR SED of SN~2004dj is plotted (a vertical shift
has been applied between the data for better visibility). This figure further
illustrates the disappearance of the CO band as well as the shifting of the peak of the SED
toward longer wavelengths. The fitting of SEDs with dust models is presented in
Section~\ref{model}.

We conclude that the MIR excess flux present in all IRAC bands and also at longer wavelengths
between $\sim$ +450 - 900 days is most likely caused by thermal emission of dust particles inside the SN ejecta. 
The significant increase and subsequent decrease of the excess emission on a timescale of a few hundred days
argues against an IR-echo of the SN radiation from pre-existing CSM or ISM, because in that
case the temporal evolution of the light curve should be at least one order of magnitude slower.    

\subsection{Polarization data}\label{anal_pol}  

\begin{figure}
\begin{center}
\leavevmode
\resizebox{\hsize}{!}{\includegraphics{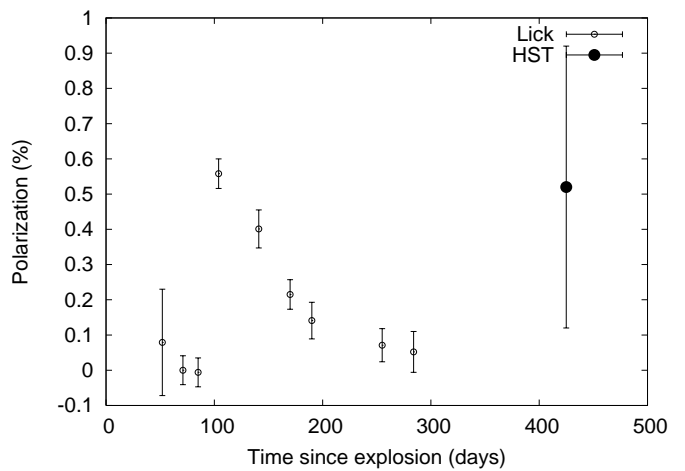}}
\caption{Evolution of the detected true degree-of-polarization in SN~2004dj. 
Open symbols: spectropolarimetry by Leonard et. al. (2006); 
filled symbol: imaging polarimetry with $HST$/ACS (this paper).}
\label{fig:pol}
\end{center}
\end{figure}

The measured true degree-of-polarization $p_{true}$ = 0.81 $\pm$ 0.4 \% 
derived in Section~\ref{obs_polar} suggests a 2-sigma (weak) detection of
polarized light from SN~2004dj in the optical (close to $B$-band) at +425 days.
This phase is not covered well by the {\it Spitzer} observations, but 
it is at the beginning of the "MIR bump"-phase, when the suspected
dust formation just started. The detected polarization, if real, 
fits nicely into this picture, as the photons scattered off the
dust grains are expected to be polarized. 

The source of the detected polarization may be interstellar,
not related to the SN ejecta. To prove or reject this hypothesis, we used the 
spectropolarimetric results by Leonard et al. (2006). 
They detected changing net continuum polarization of SN~2004dj
during early phases, starting close to zero, then climbing up to 
$p = 0.55$ \% at +90 days then declining as $\sim 1 / t^2$ to 0.052 \% by +270 days 
after explosion (see Fig.\ref{fig:pol}). On the other hand, the
measured polarization in the optical spectral lines was much less, going
below 0.1 \% in strong lines ($H\alpha$ or the CaII triplet) and in the 
metallic line blends shortward of 5500 \AA.

Leonard et al. (2006) also measured the interstellar polarization in the direction
of SN~2004dj, and found $p_{ISM} = 0.29$ \%. 
Obviously our detection from {\it HST} at +425 days 
is significantly higher than $p_{ISM}$ found by Leonard
et al. Subtracting $p_{ISM} \sim 0.3$ \% from the
total measured polarization, $p_{SN} \sim 0.5$ \% remains, which is
similar to the amount of net continuum polarization detected by Leonard et al. (2006)
during the photospheric phase. 

It must be emphasized that the
source of the polarization during the photospheric phase (measured by Leonard et al.
2006) and the nebular phase presented here is very different. During the
photospheric phase, the polarization is caused by Thompson-scattering, 
and the detected net polarization suggests an asymmetric shape of the inner ejecta
(Leonard et al. 2006; Wang \& Wheeler 2008). Leonard et al. also pointed out that this kind of polarization
is expected to quickly decrease with time when the ejecta dilutes and the scattering
optical depth declines, and this is exactly what they have observed (cf. Fig.\ref{fig:pol}). 
On the other hand, the polarization detected with $HST$ at +425 days,
when the ejecta are thought to be almost fully transparent and the electron 
scattering optical depth is much smaller, should be produced by scattering on other particles. 
Newly formed dust seems to be a plausible explanation, although some kind of deviation
from spherical symmetry is still necessary to produce observable net polarization.
One possibility may be the formation of dust clumps distributed asymmetrically
in the SN ejecta. This was also proposed by e.g. Tran et al. (1997) to explain the
observed polarization in SN~1993J.
We conclude that the detected polarization, $p_{SN} \sim 0.5$ \% at +425 days is likely 
caused by scattering on freshly formed dust particles, 
which agrees with other signs of dust formation after this epoch.

\section{Models for dust}\label{model}

In the previous section we presented several pieces of evidence for the presence of dust around SN~2004dj.
In order to analyze the physical properties and estimate the total amount of dust, we fit theoretical SEDs 
from analytic and numerical models to the measured IRAC and MIPS data 
(days 267-275, 849-883, 1006-1016 and 1236-1246, respectively). First, we use the analytic model described by 
Meikle et al. (2007), then we apply the numerical radiative-transfer code MOCASSIN (Ercolano et al. 2003, 2005).
Prior {\bf to} fitting, the observed SED fluxes were dereddened using the galactic reddening law parametrized
by Fitzpatrick \& Massa (2007) assuming $R_V = 3.1$ and adopting $E(B-V)=0.1$ from Paper~I. The IRAC data
were also corrected for the fluxes of the surrounding cluster as explained in Section~\ref{anal_s96}.
We assumed a general 10 \% uncertainty for all observed fluxes to represent the overall random plus 
systematic errors (see e.g. Kotak et al. 2005).  

\subsection{Analytic models}\label{model_an}

We fitted simple analytic models to the observed MIR SEDs using Eq.1 in Meikle et al. (2007)
assuming {\bf a} homogeneous (constant-density) dust distribution.   
To estimate the dust optical depth ($\tau_{\lambda}$), we adopted the power-law grain size distribution
of Mathis, Rumpl \& Nordsieck (1977, hereafter MRN) assuming $m$ = 3.5 for the power-law index and 
$a_{min}$ = 0.005 $\mu$m and $a_{max}$ = 0.05 $\mu$m for the minimum and maximum grain sizes, respectively. 

As there is no detectable 9.7 micron silicate feature  in any of the observed SEDs, our initial fits using 
C-Si-PAH dust composition (Weingartner \& Draine 2001) failed. 
The lack of detectable SiO emission in the IRS spectra (Fig.\ref{fig:irs}) also supports the absence of silicates
in the dust composition.
 
Thus, we chose amorphous carbon (AC) grains for subsequent models. Values for the dust opacity 
$\kappa_{\lambda}$ were taken from Colangeli et al. (1995).
For the grain density we adopted $\rho_{gr} = 1.85$ g cm$^{-3}$ as the value for the AC1 
material given by Rouleau \& Martin (1991) (see e.g. Kotak et al. 2009; Botticella et al. 2009).
The grain temperature $T$ and the grain number density scaling factor ($k$) were free parameters during the fitting.
The dust is assumed to be distributed uniformly within a sphere, the radius $R$ of which was chosen using the maximum 
velocity of the SN ejecta during the nebular phase (v$_{max} \sim$ 3250 km s$^{-1}$, Paper II; 
see the method in Meikle et al. 2007), and assuming homologous expansion, 
i.e. $R ~=~ v_{max} \cdot t$ where $t$ is the phase of the SN at the moment of the 
SED observation. 

For SN~1987A Wooden et al. (1993) showed that a hot component ($T \sim 5000$ K, probably caused by 
an optically thick gas within the ejecta) affects the dust-emission continuum. 
On the other hand, Meikle et al. (2007) and Kotak et al. (2009) found that 
the contribution of the hot component is very small at wavelengths 
longer than 3 $\mu$m for SNe~2003gd and 2004et (also Type II-P events). 
In a $T = 6000$ K, $v_{hot} = 60$ km s$^{-1}$ blackbody at +267 days 
(Kotak et al. 2009) this contribution is less than 1\% .
Late-time optical photometry of SN~2004dj was dominated by the flux from the 
cluster Sandage-96, which
prevented the inclusion of optical measurements in the fitted SEDs. 
For these reasons, we neglected the presence of the hot component in the models. 

\begin{figure*}
\begin{center}
\includegraphics[width=8cm]{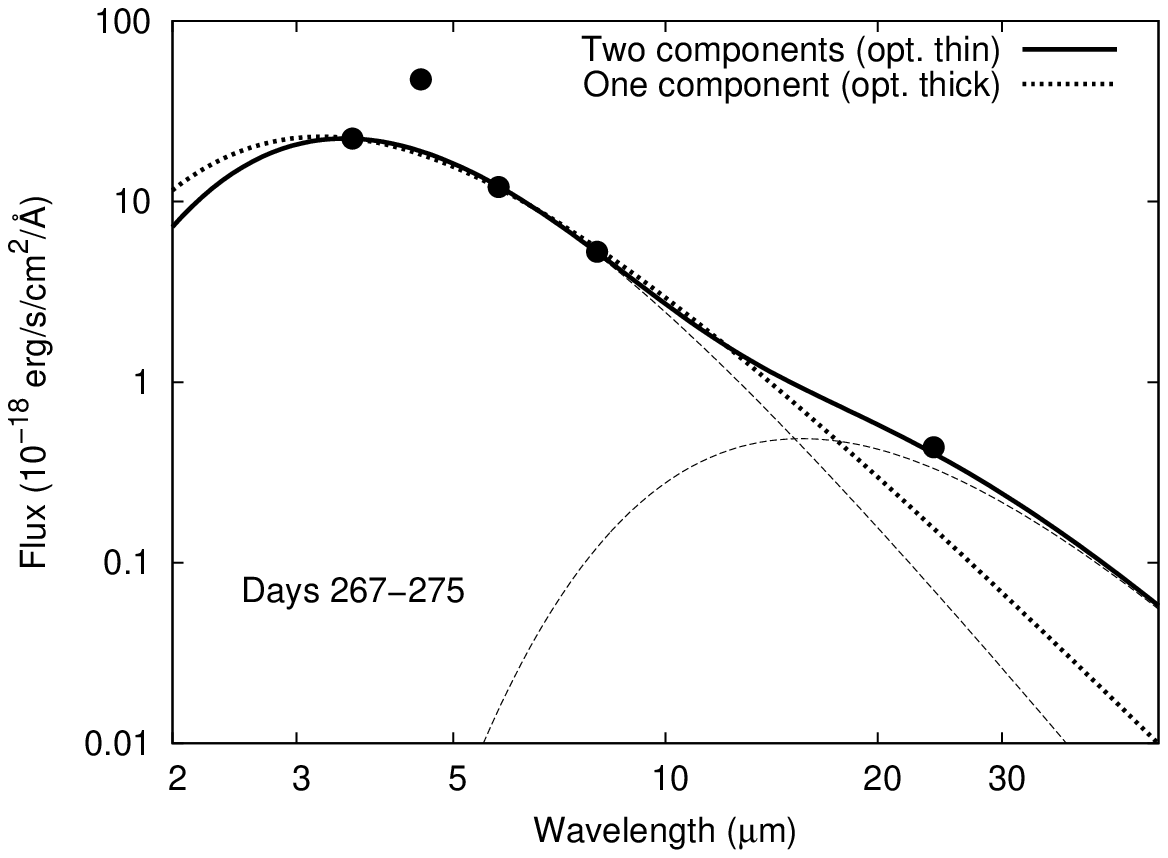}
\includegraphics[width=8cm]{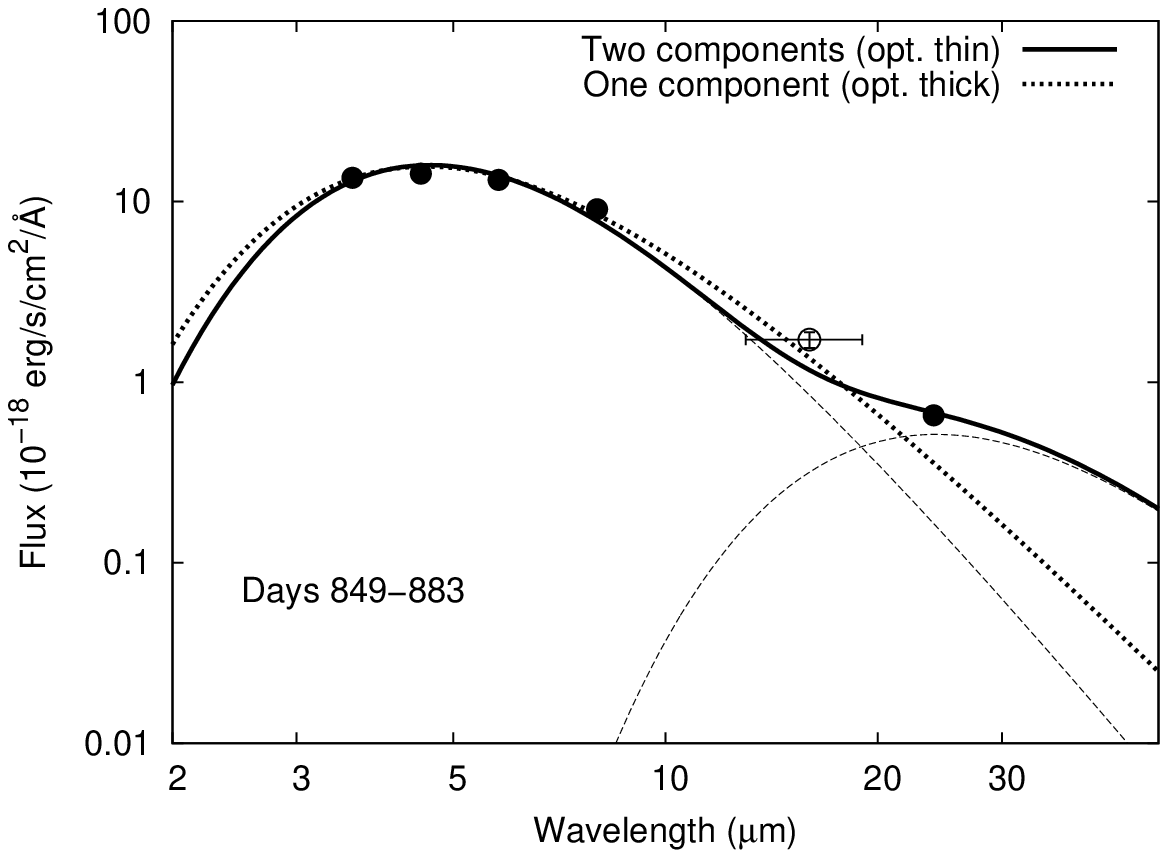}
\includegraphics[width=8cm]{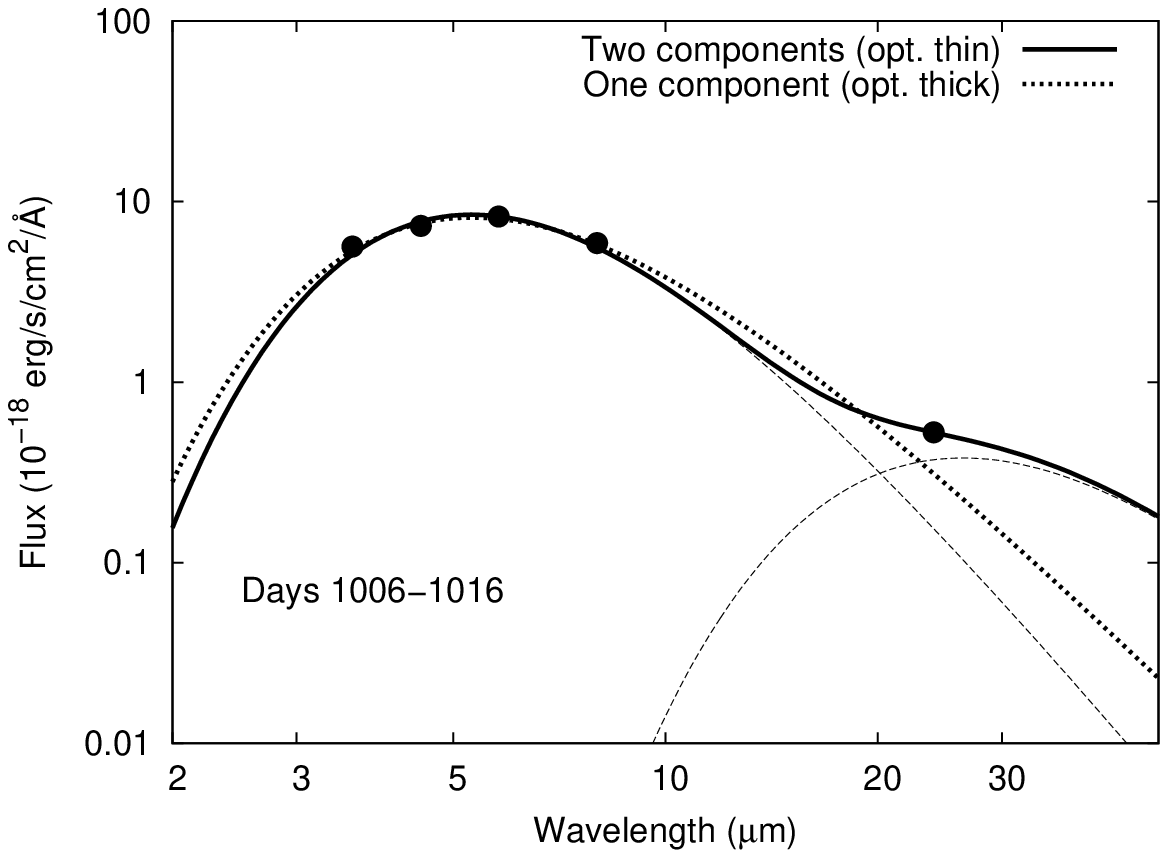}
\includegraphics[width=8cm]{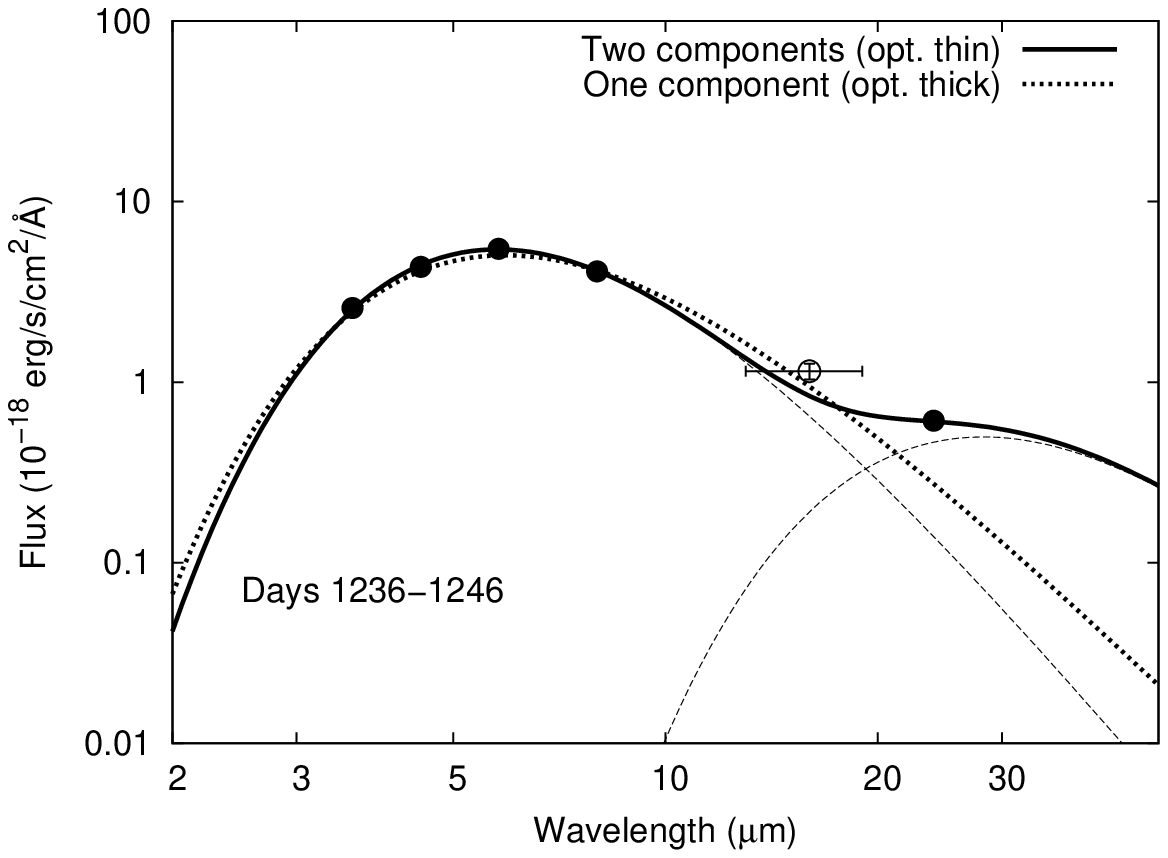}
\end{center}
\caption{Single-component blackbodies (optically thick case, dotted lines) and 
two-component analytic dust models (optically
thin case, solid lines) compared to the MIR SEDs. 
At the first epoch, we eliminated the 4.5 $\mu$m point from the fitting, because of the excess flux from the 
1-0 vibrational band of CO (see text for details). The IRS PUI fluxes are also shown as empty circles 
for illustration, 
but those were not included in the fitting.}
\label{fig:analdust}
\end{figure*}

The results are plotted in Fig. \ref{fig:analdust}. Evidently a single component, 
whether blackbody or optically thin, cannot give an adequate fit simultaneously to the IRAC and MIPS fluxes. 
All single-temperature models
underestimate the flux at 24 $\mu$m. Thus, we added a cold component to our models, similar 
to Kotak et al. (2009), which resulted in a reasonably good fit to all observed data.
However, a single flux value obtained from broadband photometry may be misleading, because it 
may not be entirely caused by pure continuum emission by a blackbody. Since none of the IRS spectra 
extend to this wavelength regime, we could not rule out that the MIPS flux is contaminated 
by line emission. On the other hand, Kotak et al. (2009) presented nebular-phase MIR
spectra of the Type II-P SN~2004et extending to 30 $\mu$m that showed no significant
emission lines around 24 $\mu$m. Assuming that SNe 2004dj and 2004et had similar 
MIR spectra we considered the MIPS fluxes as caused by pure continuum emission.

The parameters of the best-fitting models are collected in Table~\ref{tab:analpar}. 
The radius of the dust-forming zone containing the newly formed warm dust, $R_{warm}$, 
increased from 0.75 to 3.88 $\times$ 10$^{16}$ cm between 270 and 1240 days, 
while during the same time its temperature decreased from 710 to 424 K. 
The flux from the cold component can be described by a simple blackbody 
with $T_{cold}$ = 186 - 103 K and $R_{cold}$ = 1.5 - 6.2 $\times$ 10$^{16}$ cm. 

The cumulative mass of freshly formed dust was calculated as $M_d = 4 \pi R^2 \tau_{\nu} / 3 \kappa_{\nu }$ 
(Lucy et al. 1989, Meikle et al. 2007) for each fitted epoch. 
The resulting masses are between 3.1 $\times$ 10$^{-6}$ and 1.4 $\times$ 10$^{-5} M_{\odot}$, 
but it should be noted that these masses from analytic models are actually lower limits (Meikle et al. 2007),
because of the implicit assumption that the dust cloud has the lowest possible optical depth.
Theoretical studies by Kozasa et al. (2009) also suggested that assuming optically thin dust opacity 
in the MIR 
leads to an underestimation of the total dust mass. 
More realistic dust masses can be obtained only from numerical models for the optically thick dust 
cloud (see below).

The last three columns in Table~\ref{tab:analpar} contain the blackbody MIR luminosities, $L_{warm}$ and $L_{cold}$, 
calculated from the radii and temperatures of the warm and cold components, respectively. For comparison, 
the luminosity from the radioactive $^{56}$Co-decay ($L_{Co}$) is also given, assuming 0.02 $M_\odot$
for the initial Ni-mass (Paper~I). Apparently the warm component dominates the MIR luminosity, which
is comparable to $L_{Co}$ at $\sim 270$ days, but becomes orders of magnitude higher than $L_{Co}$ by
$\sim 850$ days.   

\begin{table*}
\begin{center}
\caption{\label{tab:analpar} Parameters for the best-fit analytic models to SN 2004dj SEDs. $R_{bb}$ and $T_{bb}$ belongs to the single-component blackbody,
while the other parameters correspond to the two-component analytic model (see Section 4.1).}
\newcommand\T{\rule{0pt}{3.1ex}}
\newcommand\B{\rule[-1.7ex]{0pt}{0pt}}
\begin{tabular}{c|ccccccc|ccc}
\hline
\hline
Epoch & $T_{bb}$ & $R_{bb}$ & $T_{warm}$ & $R_{warm}$ & $T_{cold}$ & $R_{cold}$ & $M_{dust}$ & $L_{warm}$ & $L_{cold}$ & $L_{Co}$ \T \\
(days) & (K) & (10$^{16}$ cm) & (K) & (10$^{16}$ cm) & (K) & (10$^{16}$ cm) & (10$^{-5} M_{\odot}$) & \multicolumn{3}{c} {($10^{38}$ erg s$^{-1}$)} \B \\
\hline
267-275 & 893 & 0.19 & 710 & 0.75 & 186 & 1.5 & 0.31 & 19.0 & 1.9 & 237\T \\		
849-883 & 634 & 0.37 & 530 & 2.48 & 120 & 4.3 & 1.11 & 18.3 & 2.7 & 1.1 \\	
1006-1016 & 545 & 0.39 & 462 & 2.85 & 110 & 4.6 & 1.32 & 11.1 & 2.2 & 0.3 \\ 	
1236-1246 & 490 & 0.40 & 424 & 3.88 & 103 & 6.2 & 1.39 & 7.8 & 3.1 & 0.04 \B \\		
\hline
\end{tabular}
\tablefoot{
Typical error is $\pm 15$ K for temperatures and $\pm 5$ $\times$ 10$^{14}$ cm for radii. $R_{warm}$ was calculated
from $v = 3250$ km s$^{-1}$ for every epoch (see text).
}
\end{center}
\end{table*}

\subsection{Numerical models}\label{model_num}

To model the optically-thick dust cloud around SN~2004dj, we applied the three-dimensional Monte Carlo 
radiative-transfer code MOCASSIN. This code was originally developed for modeling the physical conditions in
photoionized regions (Ercolano et al. 2003, 2005). The code uses a ray-tracing technique, following the paths of 
photons emitted from a given source through a spherical shell containing a specified medium. 
For numerical calculations, the shell is mapped onto a Cartesian grid allowing light-matter 
interactions (absorption, re-emission and scattering events) and to track energy packets until 
they leave the shell
and contribute to the observed SED.
In newer versions it is possible to model environments that contain not only gas but also dust, 
or to assume pure dust 
regions (Ercolano et al. 2005, 2007). This can be applied to reconstruct dust-enriched 
environments of CC SNe and to 
determine physical parameters of the circumstellar medium (grain-size distribution, 
composition and geometry), as was 
shown by Sugerman et al. (2006) and Ercolano et al. (2007).

We used version 2.02.55 of MOCASSIN for the present study. We chose amorphous carbon 
for grain composition (optical
constants were taken from Hanner, 1988) for the same reason as during the fitting of 
analytic models described above. 
The original grid in the code allowed us to model uniform spatial distribution of grains. 
Thus, we first created constant-density dust shells. 

We adopted four different grain-size distributions: MRN (with the same parameters as before) 
and three other cases of single-sized grains with radii of 0.005, 0.05 and 0.1 $\mu$m. 
With only small ($r$=0.005 $\mu$m) grains, we could not reproduce the observed SEDs, 
but we found an adequate fitting for the other three cases. It agrees well with the
calculations of Kozasa et al. (2009) that dust mass is dominated by grains with radii larger 
than 0.03 $\mu$m in Type II-P SNe. 

The final results of these calculations are shown in Table \ref{tab:numpar1}. 
For the best-fitting models, the illuminating source was given as a blackbody with $T_{BB}$ = 7000 K and 
$L_*$ = 2.2-4.5 $\times$ 10$^5 L_{\odot}$. The geometry of the dust cloud was assumed as a spherical
shell with inner and outer radii $R_{in}$ and $R_{out}$. For $R_{out}$ we adopted the radius of 
the warm component
($R_{warm}$) from the analytic models (corresponding to $v$ $\sim 3250$ km s$^{-1}$ in velocity space).
The inner radius $R_{in}$ was a free parameter, and the values that produced the best fit are shown
in Table~\ref{tab:numpar1}. 

For the dust masses we obtained values between 2-8 $\times$ 10$^{-4} M_{\odot}$, 
depending on the size and thickness of the dust cloud, 
which are more than one order of magnitude higher than the masses from analytic models. 
It should be noted, however, that in order to get 
satisfactory fitting of the MIPS fluxes, a cold blackbody component was added to
the model with the same parameters as described in Section~\ref{model_an}.  

\begin{table*}
\begin{center}
\caption{\label{tab:numpar1} Parameters for best-fit homogeneous MOCASSIN models to SN 2004dj SEDs}
\newcommand\T{\rule{0pt}{3.1ex}}
\newcommand\B{\rule[-1.7ex]{0pt}{0pt}}
\begin{tabular}{ccccc}
\hline
\hline
~ & Days 267-275 & Days 849-883 & Days 1006-1016 & Days 1236-1246 \T \B \\
\hline
$L_*$(10$^5 L_{\odot}$) & 4.5 & 4.5 & 2.8 & 2.2\T \\
$T_{BB}$(K) & 7000 & 7000 & 7000 & 7000 \\
$R_{in}$(10$^{15}$ cm) & 3.5 & 5.0 & 5.0 & 8.0 \\
$R_{out}$/$R_{in}$ & 2.1 & 5.0 & 5.7 & 4.9\B \\
\hline
Grain size: MRN & ~ & ~ & ~ & ~ \T \B \\
\hline			
$n_{dust}$(10$^{-6}$ cm$^{-3}$) & 10.0 & 10.0 & 10.0 & 2.0 \T \\
$M_{dust}$(10$^{-4} M_{\odot}$) & 0.2 & 2.0 & 4.8 & 2.6 \B \\		
\hline
Grain size: 0.05 $\mu$m & ~ & ~ & ~ & ~ \T \B \\
\hline
$n_{dust}$(10$^{-6}$ cm$^{-3}$) & 10.0 & 10.0 & 10.0 & 1.0 \T \\
$M_{dust}$(10$^{-4} M_{\odot}$) & 0.2 & 3.2 & 7.6 & 3.6 \B \\		
\hline
Grain size: 0.1 $\mu$m & ~ & ~ & ~ & ~ \T \B \\
\hline
$n_{dust}$(10$^{-6}$ cm$^{-3}$) & 2.0 & 1.0 & 1.0 & 0.3 \T \\
$M_{dust}$(10$^{-4} M_{\odot}$) & 0.2 & 2.6 & 6.2 & 4.2 \B \\		
\hline
\end{tabular}
\end{center}
\end{table*}

\begin{figure}
\begin{center}
\resizebox{\hsize}{!}{\includegraphics{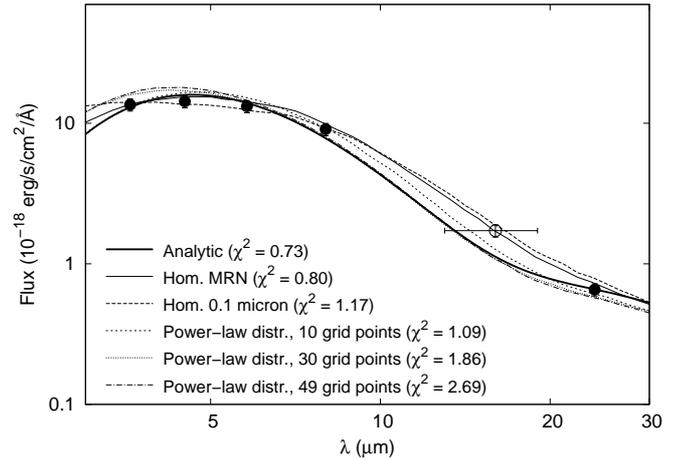}}
\caption{Comparison of the best fitting analytic and MOCASSIN models with the observed SED for 849-883 days. 
See text and Tables~\ref{tab:analpar}, \ref{tab:numpar1}, and \ref{tab:numpar2} for the explanation of
the different models. Models assuming the MRN grain-size distribution were found to be the best to reproduce the observations.
The IRS PUI flux (empty circle) was not used during the fitting.}
\label{fig:week124all}
\end{center}
\end{figure}

As a second step, we kept the MRN distribution, but generated shells with 
$\rho \propto r^{-n}$ density profiles. We chose $n=7$ because this density distribution was found
when modeling the optical spectra of SN~2004dj during the photospheric phase (Paper~I).  
These steep density distributions in the SN ejecta are also predicted by numerical 
simulations (e.g. Chugai et al, 2007; Utrobin, 2007). 
We computed symmetric grids for the density distribution with 10 grid points on each axis, 
but for comparison we 
also produced models with higher resolution grids using 15, 30 and 49 grid points. 
These models were applied for the SED observed between 849-883 days 
(see the different models on Fig.~\ref{fig:week124all}). 
The previous values for $R_{out}$ were again used.   

\begin{table*}
\begin{center}
\caption{\label{tab:numpar2} Parameters for MOCASSIN models with power-law density distribution}
\newcommand\T{\rule{0pt}{3.1ex}}
\newcommand\B{\rule[-1.7ex]{0pt}{0pt}}
\begin{tabular}{cccccc|c}
\hline
\hline
~ & $L_*$ & $R_{in}$ & $R_{out}$/$R_{in}$ & $T_{BB}$ & $n_{dust}$ & $M_{dust}$ \T \\
~ & (10$^5 L_{\odot}$) & (10$^{15}$ cm) & ~ & (K) & (10$^{-8}$ cm$^{-3}$) & (10$^{-4} M_{\odot}$) \B \\
\hline
Days 267-275 & \multicolumn{6}{c}{~} \T \B\\
\hline
10 grid points & 4.5 & 1.0 & 7.5 & 7000 & 0.2 & 0.1\T \B \\ 
\hline
Days 849-883 & \multicolumn{6}{c}{~} \T \B \\
\hline
10 grid points & 4.0 & 1.0 & 25.0 & 7000 & 1.0 & 0.4\T \\
10 grid points & 6.0 & 5.0 & 5.0 & 7000 & 10.0 & 1.3 \\
(n=2) & & & & & & \\
15 grid points & 4.0 & 1.0 & 25.0 & 7000 & 3.0 & 0.4 \\
30 grid points & 4.0 & 2.5 & 10.0 & 7000 & 800.0 & 0.5 \\
49 grid points & 4.0 & 2.0 & 12.5 & 7000 & 300.0 & 0.4\B \\
\hline
Days 1006-1016 & \multicolumn{6}{c}{~} \T \B\\
\hline
10 grid points & 2.4 & 1.0 & 28.5 & 7000 & 1.0 & 0.4\T \B \\
\hline
Days 1236-1246 & \multicolumn{6}{c}{~} \T \B \\
\hline
10 grid points & 1.5 & 1.0 & 38.8 & 7000 & 1.0 & 0.4\T \B \\
\hline
\end{tabular}
\tablefoot{
$n_{dust}$ stands for the average number density of grains. All but one model were computed
assuming n$=$7 power-law index in the density profile. The second row at days 849-883 shows 
the best-fit parameters derived assuming n$=$2 (see text for details).
}
\end{center}
\end{table*}

We show in Table~\ref{tab:numpar2} that the models with a power-law grain-density 
distribution led to somewhat different solutions, 
which resulted in dust masses one order of magnitude lower than in the case of constant-density models 
(but still 3-4 times higher than the results from analytic models).   
In order to test the effect of choosing n$=$7 on the derived dust masses,
we also generated a model assuming $\rho \propto r^{-2}$ density profile (e.g. Ercolano et al. 
2007, Wesson et al. 2010). This density distribution is expected in a CSM 
produced by a stellar wind with constant outflow velocity. The resulting dust mass
turned out to be a factor of 2 higher than in the case of n$=$7 
(see Table~\ref{tab:numpar2}), but lower than the results from the constant-density
models (Table~\ref{tab:numpar1}).

\subsection{Discussion of the dust models}\label{anal_conc}

The results of analytic and numerical modeling presented above confirm that the MIR SEDs can be well explained
by assuming newly-formed dust in the ejecta of SN 2004dj. The lack of the observed spectral features of SiO and 
tests of different models yield amorphous carbon as the most likely grain composition. The
best-fitting models give large ($a \sim$ 0.05-0.1 $\mu$m) grains, which is consistent with 
recent theoretical studies of dust in Type II-P SNe. Between +267 and +1246 days, the dust temperature decreased 
from $\sim$700 to $\sim 400$ K. The lower limit for dust mass is 1.4 $\times$ 10$^{-5}$ $M_{\odot}$, while the highest
mass derived from our models is 7.6 $\times$ 10$^{-4}$ $M_{\odot}$. 
The estimated amount of new dust around SN 2004dj is similar to those obtained for other Type II SNe, 
and therefore does not prove that CC SNe are significant dust sources in the Universe. 

Note that if the dust had been assumed to form optically-thick clumps Lucy et al. 1989), 
the resulting dust mass would have been somewhat higher. Because the mass hidden by clumping 
depends on the actual number and optical depth of individual clumps, it is difficult to predict
the dust mass unambiguously in such a model. Numerical simulations by Sugerman et al. (2006) 
and Ercolano et al. (2007) showed that the mass of hidden dust could be one order of 
magnitude higher than calculated with smooth density distribution.
We did not investigate these models in detail, but mention that even if the dust distribution were
clumpy around SN~2004dj, and the dust mass were one order of magnitude higher than estimated above,
it still would not reach 0.01 $M_\odot$.

The modeling of the SEDs revealed a cold component as a $T \sim 110$ K temperature 
shell lying at 1.5 - 6 $\times$ 10$^{16}$ cm away from the central source. 
The location of this component is in the outer region
($v \sim 6400$ km s$^{-1}$) of the SN ejecta, where the ejecta/CSM interaction is expected to take place. 
Such a cold component with similar temperature ($T \sim 100$ K) was found in SN~2004et by Kotak et al. (2009),
and also in SN~2002hh, although the latter is slightly warmer ($T \sim 300$ K, Meikle et al. 2006). 
However, the cold components in these Type II-P SNe were attributed to an IR-echo, i.e. a radiation from 
pre-existing dust reheated by the strong UV/optical radiation from the SN close to maximum light. 

\begin{figure}
\begin{center}
\includegraphics[width=7cm]{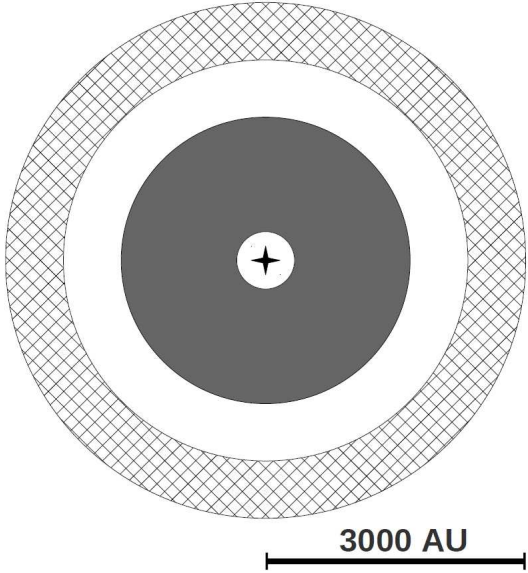}
\end{center}
\caption{Geometrical model of warm (inner, gray) and cold (outer, hatched) dust shells around SN~2004dj at
$\sim 850$ days}
\label{fig:geometry}
\end{figure}

Although pre-existing dust around SN~2004dj seems to be a plausible explanation for the early radio/X-ray 
detection (Chevalier et al. 2006), we do not favor the IR-echo hypothesis as the cause of the MIR cold component in this case. 
There are several arguments that do not support the probability of the IR-echo. First, both SNe 2002hh and 2004et were moderately
or highly reddened ($A_V > 1$ mag), while the reddening of SN~2004dj was much less ($A_V \leq 0.3$ mag, Paper~I).
Thus, pre-existing CSM dust, if present, should have been much less dense around 2004dj than in 2002hh or 2004et.
Second, the surrounding cluster, S96, is a young ($t \sim 10$ Myr), very compact cluster with a significant
OB-star population (Maiz-Apellaniz et al. 2004, Paper~II). Strong UV/optical radiation from nearby OB-stars
should expel the ISM/dust from the cluster, although some CSM around massive stars resulting from mass loss via 
stellar wind might still be present. Third, the early UV/X-ray flash of the SN should create a dust-free cavity within a region of $\sim 10^{16}$ - $10^{17}$ 
cm in size after core collapse and around maximum light (Dwek, 1983, 1985).  
The radii of the dust models that fit the observed MIR SEDs are within this region, therefore the MIR radiation
should come mostly from inside the cavity. Pre-existing dust within this region is unlikely.   
We also emphasize that the dust cloud around SN~2004et was found to have a roughly constant
size (Kotak et al. 2009), while our models for SN~2004dj clearly indicate that the dust regions expand
homologuously with the ejecta. For these reasons, we do not expect that either the cold or the warm 
component is caused by IR-echo, although some pre-existing CSM at higher
distance cannot be ruled out completely. 

Another more probable possibility for the cold component is the condensation of dust grains in a CDS 
between the forward and reverse shocks (Section~\ref{intro}). 
Although this mechanism is supposed to work predominantly in Type IIn SNe because of the higher density 
CSM, there are hints of a similar process taking place around SN 2004dj.  
Chevalier et al. (2006) pointed out based on radio and X-ray observations of Beswick et al. (2005) and 
Pooley \& Lewin (2004) that there is a non-negligible amount of CSM around SN~2004dj, probably produced by stellar wind from 
the progenitor, which has an estimated mass-loss rate of $\sim$10$^{-6}$ M$_{\odot}$ yr$^{-1}$. 
Chugai et al. (2007) presented similar results from studying the observed H$\alpha$ profiles during the photospheric and
early nebular phase. They argued that the observed high-velocity absorption 'notch' features 
could not form in the unshocked ejecta, instead, they are supposed to be produced by the CDS modified by Rayleigh-Taylor instability.
Applying Eq.10 of Chugai et al. (2007), the expected radius of the continously growing CDS is 
1.9 - 7.0 $\times$ 10$^{16}$ cm between +267 and +1246 days in a self-similar model. 
These values agree well with the size of the cold component involved in our models (see Table \ref{tab:analpar}), which
strengthens the assumption that CDS plays a role in producing the MIR SEDs.

\section{Summary}\label{sum}

Using public data of {\it Spitzer-} and {\it Hubble Space Telescope}, we presented a detailed analysis of MIR light curves and
spectra on SN 2004dj between +98 and +1381 days after explosion. Following SN 2004et (Kotak et al. 2009), 
this is the second long-term
study of the dust-formation processes around a Type II-P supernova. 
We found several pieces of evidence for dust formation after explosion. These include 
\begin{itemize}
\item{significant brightening in MIR light curves starting after +400 days},
\item{detection of $\sim 0.5$ \% polarization from the SN ejecta in the optical ({\it HST F435W} filter) at +425 days.}
\end{itemize}

We fitted analytic as well as numerical models to the observed MIR SEDs of SN~2004dj by applying the model 
by Meikle et al. (2007) and
the 3D radiative-transfer code MOCASSIN. The models confirmed the presence of dust as early as 
$\sim$ +270 days and showed that
the most intensive period of dust formation occured between $\sim$ +270 and 1000 days.
We found that the observed SEDs require the presence of a "warm" ($T \sim 500$ K)
and a "cold" ($T \sim 100$ K) dust component. The "warm" component probably consists of freshly-formed amorphous carbon grains
inside the SN ejecta at $v$ $\sim 3200$ km s$^{-1}$. The "cold" component is located at $v$ $\sim 6400$ km s$^{-1}$,
which is close to the region between the forward and reverse shock, where the cool dense shell is expected to be formed 
during the ejecta-CSM interaction (Chugai et al. 2007). Using smooth dust density distributions, the minimum dust mass was 
found to be between $\sim 10^{-5}$ - 10$^{-4}$ $M_{\odot}$ depending on the applied model, but this could be one 
order of magnitude higher, $\sim 10^{-3}$ $M_{\odot}$, if the dust distribution were clumpy. However, this is still 
several orders of magnitude less than what is needed to explain the significant dust content in star-forming galaxies.

\section*{Acknowledgments} 

We would like to thank the referee J. Danziger for the critical but very useful comments, which
helped us to improve the paper. Thanks are also due to P. Meikle for his valuable comments and suggestions on the dust
models, B. Ercolano for sending us her MOCASSIN code ver. 2.02.55 and for her extensive help in running the code, and L. 
Colangeli and V. Mennella for providing the electronic version of their table about mass-extinction coefficients of 
carbon grains. Fruitful discussions with N. Chugai, J. C. Wheeler, and S. D. Van Dyk are gratefully acknowledged.  
This work is based on observations made with the {\it Spitzer Space Telescope}, which is operated by the
Jet Propulsion Laboratory, California Institute of Technology under a contract with NASA.
Some of the data presented in this paper were obtained from the Multimission Archive at the Space Telescope 
Science Institute (MAST). STScI is operated by the Association of Universities for Research in Astronomy, 
Inc., under NASA contract NAS5-26555. Support for MAST for non-HST data is provided by the NASA Office of 
Space Science via grant NNX09AF08G and by other grants and contracts.
This work is supported by the Hungarian OTKA Grants K76816 and MB08C 81013, the University of Sydney, 
NSF Grant AST-0707669, the Texas
Advanced Research Program grant ASTRO-ARP-0094 and the "Lend\"ulet" Young Researchers'
Program of the Hungarian Academy of Sciences.


\begin{thebibliography}{}

\bibitem[]{}
  Andrews, J. E., et al. 2010, ApJ, 715, 541

\bibitem[]{}
  Barlow, M. J., et al. 2005, ApJ, 627, L113

\bibitem[]{}
  Bertoldi, F., Carilli, C. L., Cox, P., Fan, X., Strauss, M. A., Beelen, A., Omont, A., \& Zylka, R. 2003, A\&A, 406, L55

\bibitem[]{}
  Beswick, R. J., Muxlow, T. W. B., Argo, M. K., Pedlar, A., Marcaide, J. M., \& Wills, K. A. 2005, ApJ, 623, L21

\bibitem[]{}
  Bianchi, S., \& Schneider, R. 2007, MNRAS, 378, 973

\bibitem[]{}
  Biretta, J., Kozhurina-Platais, V. Boffi, F., Sparks, W., \& Walsh, J., 2004, STScI Instrument Science Report ACS 2004-09

\bibitem[]{}
  Blair, W. P., Ghavamian, P., Long, K. S., Williams, B. J., Borkowski, K. J., Reynolds, S. P., \& Sankrit, R. 2007, ApJ,
  662, 998

\bibitem[]{}
  Bode, M. F., \& Evans, A. 1980, MNRAS, 193, 21

\bibitem[]{}
  Botticella, M. T., et al. 2009, MNRAS, 398, 1041

\bibitem[]{}
  Cernuschi, F., Marsicano, F. R., \& Codina, S. 1967, Ann. d'Astrophys., 30, 1039

\bibitem[]{}
  Chevalier, R. A. 1982, ApJ, 258, 790

\bibitem[]{}
  Chevalier, R. A., Fransson, C., \& Nymark, T. K. 2006, ApJ, 641, 1029

\bibitem[]{}
  Chieffi, A. \& Limongi, M. 2004, ApJ, 608, 405 

\bibitem[]{}
  Chugai, N. N., Chevalier, R. A., \& Utrobin, V. P. 2007, ApJ, 662, 1136

\bibitem[]{}
  Clayton, D. D. 1979, Ap\&SS, 65, 179

\bibitem[]{}
  Clayton, D. D., Amari, S., \& Zinner, E. 1997, Ap\&SS, 251, 355

\bibitem[]{}
  Clayton, D. D., \& Nittler, L. R. 2004, ARA\&A, 42, 39

\bibitem[]{}
  Colangeli, L., Mennella, V., Palumbo, P., Rotundi, A., \& Bussoletti, E. 1995, A\&AS, 113, 561

\bibitem[]{}
  Danziger, I. J., Gouiffes, C., Bouchet, P., \& Lucy, L. B. 1989, IAU Circ., 4746

\bibitem[]{}
  Danziger, I. J., Lucy, L. B., Bouchet, P., \& Gouiffes, C. 1991, in The Tenth Santa Cruz Workshop in Astronomy and Astrophysics, ed. S. E. Woosley 
  (New York: Springer-Verlag), 69

\bibitem[]{}
  Dolphin, A. E. 2000, PASP, 112, 1383

\bibitem[]{}
  Dunne, L., Eales, S., Ivison, R., Morgan, H. L., \& Edmunds, M. G. 2003, Nature, 424, 285

\bibitem[]{}
  Draine, B. T. 2003, ARA\&A, 41, 241
  
\bibitem[]{}
  Draine, B. T. 2009, in Cosmic Dust - Near and Far, ASP Conference Series, Vol. 414, ed. T. Henning, E. Grün, \& J. Steinacker (San Francisco: Astromomical Society of the Pacific), 453  

\bibitem[]{}
  Dwek, E. 1983, ApJ, 274, 175
  
\bibitem[]{}  
  Dwek, E. 1985, ApJ, 297, 719
 
\bibitem[]{}
  Dwek, E., Galliano, F., \& Jones, A. P. 2007, ApJ, 662, 927

\bibitem[]{}
  Elmhamdi, A., Danziger, I. J., Cappellaro, E., Della Valle, M., Gouiffes, C., Philips, M. M., \& Turatto, M. 2004, A\&A, 426, 963

\bibitem[]{}
  Elmhamdi, A., Danziger, I. J., Chugai, N., Pastorello, A., Turatto, M., Cappellaro, E., Altavilla, G., Benetti, S., Patat, F., \& Salvo, M. 2003, MNRAS, 338, 939

\bibitem[]{}
  Elvis, M., Marengo, M., \& Karovska, M. 2002, ApJ, 567, L107

\bibitem[]{}
  Engelbracht, C. W., et al. 2007, PASP, 119, 994
  
\bibitem[]{}
  Ercolano, B., Barlow, M. J., \& Storey, P. J. 2005, MNRAS, 362, 1038
  
\bibitem[]{}
  Ercolano, B., Barlow, M. J., Storey, P. J., \& Liu X.-W. 2003, MNRAS, 340, 1153
  
\bibitem[]{}
  Ercolano, B., Barlow, M. J., \& Sugerman, B. E. K. 2007, MNRAS, 375, 753
  
\bibitem[]{}
  Fitzpatrick, E. L., \& Massa, D. 2007, ApJ, 663, 320    

\bibitem[]{}
  Fox, O., et al. 2009, ApJ, 691, 650

\bibitem[]{}
  Gerardy, C. L., et al. 2002, ApJ, 575, 1007

\bibitem[]{}
  Hanner, M. S. 1988, NASA Conf. Publ., 3004, 22

\bibitem[]{}
  Hoyle, F., \& Wickramasinghe, N. C. 1970, Nature, 226, 2

\bibitem[]{}
  Kotak, R., Meikle, W. P. S., van Dyk, S. D., H\"oflich, P. A., \& Mattila, S. 2005, ApJ, 628, L123

\bibitem[]{}
  Kotak, R., et al. 2006, ApJ, 651, L117

\bibitem[]{}
  Kotak, R., et al. 2009, ApJ, 704, 306
  
\bibitem[]{}
  Kozasa, T., Nozawa, T., Tominaga, N., Umeda, H., Maeda, K., \& Nomoto, K. 2009, in Cosmic Dust - Near and Far, ASP Conference Series, Vol. 414, ed. T. Henning, E. Gr\"un, \& J. Steinacker (San Francisco: Astromomical Society of the Pacific), 43  

\bibitem[]{}
  Krause, O., Birkmann, S. M., Rieke, G. H., Lemke, D., Klaas, U., Hines, D. C., \& Gordon, K. D. 2004, Nature, 432, 596

\bibitem[]{}
  Leonard, D. C. et al. 2006, Nature, 440, 505

\bibitem[]{}
  Lucy, L. B., Danziger, I. J., Gouiffes, C., \& Bouchet, P. 1989, in Structure and Dynamics of the Interstellar Medium, ed.
  G. Tenorio-Tagle et al. (Berlin: Springer), 164

\bibitem[]{}
  Lucy, L. B., Danziger, I. J., Gouiffes, C., \& Bouchet, P. 1991, in The Tenth Santa Cruz Workshop in Astronomy and Astrophysics, ed. S. E. Woosley 
  (New York: Springer-Verlag), 82

\bibitem[]{}
  Maiolino, R., Schneider, R., Oliva, E., Bianchi, S., Ferrara, A., Mannucci, F., Pedani, M., \& Roca Sogorb, M. 2004,
  Nature, 431, 533

\bibitem[]{}
Ma{\'{\i}}z-Apell{\'a}niz, J., Bond, H.~E., Siegel, M.~H., Lipkin, Y.,
Maoz, D., Ofek, E.~O., \& Poznanski, D.\ 2004, \apjl, 615, L113


\bibitem[]{}
  Markwick-Kemper, F., Gallagher, S. C., Hines, D. C., \& Bouwman, J. 2007, ApJ, 668, L107

\bibitem[]{}
  Mathis, J. S., Rumpl, W., \& Nordsieck, K. H. 1977, ApJ, 217, 425
  
\bibitem[]{}
  Mattila, S., et al. 2008, MNRAS, 389, 141

\bibitem[]{}
  Matsuura, M., et al. 2009, MNRAS, 396, 918

\bibitem[]{}
  Meikle, W. P. S., et al. 2006, ApJ, 649, 332

\bibitem[]{}
  Meikle, W. P. S., et al. 2007, ApJ, 665, 608

\bibitem[]{}
  Michalowski, M. J., Watson, D., \& Hjorth, J. 2010, ApJ, 712, 942

\bibitem[]{}
  Morgan, H. L., Dunne, L., Eales, S. A., Ivison, R. J., \& Edmunds, M. G. 2003, ApJ, 597, L33

\bibitem[]{}
  Morgan, H. L., \& Edmunds, M. G. 2003, MNRAS, 343, 427

\bibitem[]{}
  Nakano, S., Itagaki, K., Bouma, R. J., Lehky, M., \& Homoch, K. 2004, IAU Circ., 8377

\bibitem[]{}
 Nomoto, K. et al. 2006, Nuclear Physics A, 777, 424

\bibitem[]{}
  Nozawa, T., et al. 2008, ApJ, 684, 1343

\bibitem[]{}
  Nozawa, T., Kozasa, T., Umeda, H., Maeda, K., \& Nomoto, K. 2003, ApJ, 598, 785

\bibitem[]{}
  Patat, F., Benetti, S., Pastorello, A., \& Filippenko, A. V. 2004, IAU Circ., 8378

\bibitem[]{}
  Pei, Y. C., Fall, S. M., \& Bechtold, J. 1991, ApJ, 378, 6

\bibitem[]{}
  Pettini, M., King, D. L., Smith, L. J., \& Hunstead, R. W. 1997, ApJ, 478, 536

\bibitem[]{}
  Pooley, D., \& Lewin, W. H. G. 2004, IAU Circ. 8390

\bibitem[]{}
  Pozzo, M., Meikle, W. P. S., Fassia, A., Geballe, T., Lundqvist, P., Chugai, N. N., \& Sollerman, J. 2004, MNRAS, 352, 457

\bibitem[]{}
  Prieto, J. L., Kistler, M. D., Thompson, T., Ykse, H., Kochanek, C. S., Stanek, K. Z., Beacom, J. F., Martini, P.,
  Pasquali, A., \& Bechtold, J. 2008, ApJ, 681, L9

\bibitem[]{}
  Reach, W. T., et al. 2006, Infrared Array Camera Data Handbook, ver. 3.0 (Spitzer Science Center, California Institute 
  of Technology, Pasadena, California 91125 USA)

\bibitem[]{}
  Rho, J., Kozasa, T., Reach, W. T., Smith, J. D., Rudnick, L., DeLaney, T., Ennis, J. A., Gomez, H., \& Tappe, A. 2008,
  ApJ, 673, 271

\bibitem[]{}
  Roche, P. F., Aitken, D. K., \& Smith, C. H. 1993, MNRAS, 261, 522
  
\bibitem[]{}
  Rouleau, F., \& Martin, P. G. 1991, ApJ, 377, 526

\bibitem[]{}
  Sakon, I., et al. 2009, ApJ, 692, 546

\bibitem[]{}
  Sandstrom, K. M., Bolatto, A. D., Stanimirovic, S., van Loon, J. T., \& Smith, J. D. 2009, ApJ, 696, 2138

\bibitem[]{}
  Schuster, M. T., Marengo, M., \& Patten, B. M. 2006, IRACproc: A software suite for processing and analyzing 
  Spitzer/IRAC data, in Observatory Operations: Strategies, Processes, and Systems. Edited by David R. Silva and Rodger 
  E. Doxsey. Proceedings of the SPIE, Volume 6270, p. 627020    

\bibitem[]{}
  Silvia, D. W., Smith, B. D., \& Shull, J. M. 2010, ApJ, 715, 1575

\bibitem[]{}
  Smith, N., et al. 2009, ApJ, 695, 1334

\bibitem[]{}
  Smith, N., Foley, R. J., Filippenko, A. V. 2008a, ApJ, 680, 568

\bibitem[]{}
  Smith, N., Chornock, R., Li, W., Ganeshalingam, M., Silverman, J. M., Foley, R. J., Filippenko, A. V., Barth, A. J. 2008b, ApJ, 686, 467

\bibitem[]{}
  Sparks, W. B., \& Axon, D. J. 1999, PASP, 111, 1298

\bibitem[]{}
  Stanimirovic, S., Bolatto, A. D., Sandstrom, K. M., Leroy, A. K., Simon, J. D., Gaensler, B. M., Shah, R. Y., \& Jackson, J. M. 2005, ApJ, 632, L103

\bibitem[]{}
   Stobie, E. \& Ferro, A. 2006 in Astronomical Data Analysis Software and Systems XV 
   ASP Conference Series, Vol. 351, Proceedings of the Conference Held 2-5 October 2005 
   in San Lorenzo de El Escorial, Spain. Edited by Carlos Gabriel, Christophe Arviset, 
   Daniel Ponz, and Enrique Solano. San Francisco: Astronomical Society of the Pacific, p.540

\bibitem[]{}
  Stratta, G., Maiolino, R., Fiore, F., \& D'Elia, V. 2007, ApJ, 661, L9

\bibitem[]{}
  Sugerman, B. E. K. 2003, AJ, 126, 1939

\bibitem[]{}
  Sugerman, B. E. K., et al. 2006, Science, 313, 196

\bibitem[]{}
 Thielemann, F-K, Nomoto, K., Hashimoto, M. 1996, ApJ, 460, 408

\bibitem[]{}
  Todini, P., \& Ferrara, A. 2001, MNRAS, 325, 726

\bibitem[]{}
  Tominaga, N., et al. 2008, ApJ, 687, 1208

\bibitem[]{}
  Tran, H. D. et al. 1997, PASP, 109, 489 

\bibitem[]{}
  Utrobin, V. P. 2007, A\&A, 461, 233

\bibitem[]{}
  Valiante, R., Schneider, R., Bianchi, S., \& Andersen, A. C. 2009, MNRAS, 397, 1661
  
\bibitem[]{}
  Vink\'o, J., et al. 2006, MNRAS, 369, 1780 (Paper I) 

\bibitem[]{}
  Vink\'o, J., et al. 2009, ApJ, 695, 619 (Paper II)

\bibitem[]{}
 Wang, L., \& Wheeler, J.~C.\ 2008, \araa, 46, 433 

\bibitem[]{}
  Wang, X., Yang, Y., Zhang, T., Ma, J., Zhou, X., Li, W., Lou, Y.-Q., \& Li, Z. 2005, ApJ, 626, L89

\bibitem[]{}
  Weingartner, J. C., \& Draine, B. T. 2001, ApJ, 548, 296

\bibitem[]{}
  Wesson, R., et al. 2010, MNRAS, 403, 474


\bibitem[]{}
  Wooden, D. H., Rank, D. M., Bregman, J. D., Witteborn, F. C., Tielens, A. G. G. M., Cohen, M., Pinto, P. A., \& Axelrod,
  T. S. 1993, ApJS, 88, 477
  
\bibitem[]{}
 Woosley, S. E., \& Weaver, T. A. 1995, ApJS, 101, 181

\bibitem[]{}
  Woosley, S. E., Heger, A., \& Weaver, T. A. 2002, Rev. Mod. Phys., 74, 1015

\bibitem[]{}
  Zafar, T., Watson, D. J., Malesani, D., Vreeswijk, P. M., Fynbo, J. P. U., Hjorth, J., Levan, A. J., \& Michalowski, M. J.
  2010, A\&A, 515, 94



\end{thebibliography}
\end{document}